\input harvmac
\noblackbox

\def\bfone{\relax{\rm 1\kern-.35em 1}}
\def\inbar{\vrule height1.5ex width.4pt depth0pt}

\def\IC{\relax\,\hbox{$\inbar\kern-.3em{\rm C}$}}
\def\ID{\relax{\rm I\kern-.18em D}}
\def\IF{\relax{\rm I\kern-.18em F}}
\def\IH{\relax{\rm I\kern-.18em H}}
\def\II{\relax{\rm I\kern-.17em I}}
\def\IN{\relax{\rm I\kern-.18em N}}
\def\IP{\relax{\rm I\kern-.18em P}}
\def\IQ{\relax\,\hbox{$\inbar\kern-.3em{\rm Q}$}}
\def\us#1{\underline{#1}}
\def\IR{\relax{\rm I\kern-.18em R}}
\font\cmss=cmss10 \font\cmsss=cmss10 at 7pt
\def\ZZ{\relax\ifmmode\mathchoice
{\hbox{\cmss Z\kern-.4em Z}}{\hbox{\cmss Z\kern-.4em Z}}
{\lower.9pt\hbox{\cmsss Z\kern-.4em Z}}
{\lower1.2pt\hbox{\cmsss Z\kern-.4em Z}}\else{\cmss Z\kern-.4em
Z}\fi}
  \def\d{\delta}
 \def\c{\gamma}
 
 \def\s{\sigma}

\def\nup#1({Nucl.\ Phys.\ $\us {B#1}$\ (}
\def\plt#1({Phys.\ Lett.\ $\us  {B#1}$\ (}
\def\cmp#1({Comm.\ Math.\ Phys.\ $\us  {#1}$\ (}
\def\prp#1({Phys.\ Rep.\ $\us  {#1}$\ (}
\def\prl#1({Phys.\ Rev.\ Lett.\ $\us  {#1}$\ (}
\def\prv#1({Phys.\ Rev.\ $\us  {#1}$\ (}
\def\mpl#1({Mod.\ Phys.\ Let.\ $\us  {A#1}$\ (}
\def\ijmp#1({Int.\ J.\ Mod.\ Phys.\ $\us{A#1}$\ (}
\def\jag#1({Jour.\ Alg.\ Geom.\ $\us {#1}$\ (}
\def\tit#1|{{\it #1},\ }

\def\Coe#1.#2.{{#1\over #2}}
\def\coeff#1#2{\relax{\textstyle {#1 \over #2}}\displaystyle}
\def\coe#1.#2.{\relax{\textstyle {#1 \over #2}}\displaystyle}
\def\half{{1 \over 2}}

\def\br{\hfill\break}
\def\wT{\widetilde T}
\def\lsw{\lambda_{\rm SW}}

%
%
\lref\LMW{W.~Lerche, P.~Mayr and
N.P.~Warner, {\it Non-Critical Strings, Del Pezzo Singularities
and Seiberg-Witten Curves,} CERN-TH/96-326,USC-96/026,
hep-th/9612085.}
\lref\JMDN{J.A.~Minahan and D.~Nemeschansky, \nup{464}(1996) 3,
hep-th/9507032; \nup{468} (1996) 72, hep-th/9601059.}
\lref\KMV{A.\ Klemm, P.\ Mayr and C.\ Vafa,
{\it BPS states of exceptional non-critical strings,}
CERN-TH-96-184, hep-th/9607139.}
\lref\GMS{O.\ Ganor, D.\ Morrison and N.\ Seiberg,
\nup{487} (1997) 93,  hep-th/9610251.}
\lref\JHGM{J.~Harvey and G.~Moore, \nup{463} (1996) 315,
\hep-th/9510182; {\it On the Algebras of BPS States,}
hep-th/9609017 .}
\lref\MKYuM{M.~Kontsevich and Yu.~Manin. \cmp{164} (1994) 525,
hep-th/9402147.}
\lref\claude{C.~Itzykson, \ijmp{B8} (1994) 1994.}
\lref\EnNCS{O.\ Ganor and A.\ Hanany, \nup{474} (1996) 122,
hep-th/9602120; \br
N.\ Seiberg and E.\ Witten, \nup{471} (1996) 121,
hep-th/9603003; \br
M.~Duff, H.~Lu and C.N.~Pope, \plt{378} (1996) 101,
hep-th/9603037; \br
M.R.~Douglas, S.~Katz, C.~Vafa, {\it
Small Instantons, del Pezzo Surfaces and Type I' theory,}
hep-th/9609071; \br
E.~Witten, \mpl{11} (1996) 2649, hep-th/9609159.}
\lref\OGan{O.~Ganor,  \nup{479} (1996) 197, hep-th/9607020;
\nup{488} (1997) 223, hep-th/9608109.}
\lref\MandFWit{E.~Witten, \nup{471} (1996) 195,
hep-th/9603150.}
\lref\DMPC{D.R.~Morrison,  private communication.}
\lref\NSeib{N.~Seiberg, \plt{388} (1996) 753,
hep-th/9608111; \br
D.R.~Morrison and  N.~Seiberg,  \nup{483} (1997) 229,
hep-th/9609070.}
\lref\SW{N.\ Seiberg and E.\ Witten, \nup426(1994) 19,
hep-th/9407087; \nup431(1994) 484, hep-th/9408099.}
\lref\KLMVW{A.\ Klemm, W.\ Lerche, P.\ Mayr, C.\ Vafa and
N.P.\ Warner,\nup{477} (1996) 746, hep-th/9604034.}
\lref\PMPC{P.~Mayr,  private communication.}
\lref\CVPC{C.~Vafa,  private communication.}
\lref\ABSS{A.\ Brandhuber and S.\ Stieberger,
{\it Self-dual Strings and Stability of BPS States in $N=2$ $SU(2)$
Gauge Theories,} hep-th/9610053.}
\lref\WLer{W.\ Lerche,
{\it Introduction to Seiberg-Witten Theory and its Stringy Origin},
hep-th/9611190.}
\lref\JRab{J.\ Rabin,  {\it Geodesics and BPS StatesGEODESICS
in $N=2$ Supersymmetric QCD}, hep-th/9703145. }
\lref\JSNW{J.~Schulze and N.P.~Warner, {\it BPS Geodesics in $N=2$
Supersymmetric Yang-Mills Theory,} preprint USC-97-001,
to appear in {\it Nuclear Physics B}, hep-th/9702012.}
%
%
\Title{\vbox{
\hbox{USC-97/006}
\hbox{NSF-ITP-97-055}
\hbox{\tt hep-th/9705237}
}}{Investigating the BPS Spectrum of Non-Critical $E_n$
Strings}

\centerline{J.A.~Minahan, D.~Nemeschansky}
\bigskip
\centerline{{\it Physics Department, U.S.C.}}
\centerline{{\it University Park, Los Angeles, CA 90089}}
\bigskip
\centerline{and}
\bigskip
\centerline{N.P.~Warner}
\bigskip
\centerline{{\it Institute for Theoretical Physics}}
\centerline{{\it University of California, Santa Barbara,
 CA 93106-4030 \footnote{*}{On leave from Physics Department,
U.S.C., University Park, Los Angeles, CA 90089}}}
\bigskip

\vskip .3in
We use the effective action of the $E_n$ non-critical
strings to study its BPS spectrum for $0 \le n \le 8$.  We show
how to introduce mass parameters, or Wilson lines, into the
effective action, and then perform the appropriate
asymptotic expansions that yield the BPS spectrum. The
result is the $E_n$ character expansion of the spectrum,
and is equivalent to performing the mirror map
on a Calabi-Yau with up to nine K\"ahler moduli.
This enables a much more detailed examination of the
$E_n$ structure of the theory, and provides extensive
checks on the effective action description of the non-critical
string.  We extract some universal ($E_n$ independent)
information concerning the degeneracies of BPS excitations.

\vskip .3in


\Date{\vbox{\hbox{USC-97/006 \qquad NSF-ITP-97-055}
\hbox{\sl {May, 1997}}}}

%
\parskip=4pt plus 15pt minus 1pt
\baselineskip=15pt plus 2pt minus 1pt
%
\newsec{Introduction}

The $E_n$ non-critical strings are ubiquitous in the
formulation of non-perturbative string theory, and
understanding these highly non-trivial fixed points
is becoming of increasing importance.  Even though there
are many ways of characterizing such strings \refs{\EnNCS,
\MandFWit,\OGan}, there is, as yet, no explicit action or
intrinsic formulation.  Descriptions of such non-critical strings
are either based upon classical
solutions of low energy effective actions, or involve
interpolating between branes, or wrapping
branes around vanishing cycles.
We will consider the string in its incarnation
in the type IIA theory (or $M$-theory) compactified
to four (or five)  dimensions
on an elliptically fibered Calabi-Yau manifold that is
also a $K3$ fibration. This corresponds to the (six-dimensional)
non-critical  string compactified to four (five) dimensions
on a torus (circle).  The non-critical string emerges, {\it i.e.}
has excitations of low mass, when a  $4$-cycle in the
Calabi-Yau manifold becomes extremely small
\refs{\MandFWit,\KMV}. The magnetic non-critical string is
obtained by wrapping
a five-brane around the collapsing $4$-cycle, while the
dual electric states of the non-critical string are
obtained by wrapping the membrane around $2$-cycles within
the $4$-cycle.    If the  collapsing $4$-cycle is a $B_n$ del Pezzo
surface  then the string is endowed with an $E_n$ symmetry:
the del Pezzo surface has
$(n+1)$ $2$-cycles, $n$ of which are acted upon naturally
by the Weyl group of $E_n$.  The remaining $2$-cycle is the
canonical divisor of the del Pezzo, whose K\"ahler modulus, $k_D$,
determines the scale of the del Pezzo, and hence determines
the string tension.
Associated with the non-critical string is the anti-self-dual
tensor multiplet, which in four (five) dimensions gives
rise to a vector multiplet.   The vanishing canonical
$2$-cycle in the del Pezzo provides the harmonic
$2$-form need to make the $U(1)$ gauge field strength.

This compactification of the string has two natural phases
that are separated by a flop transition, and are referred to as
phases $I$ and $II$ in \LMW.   They are defined as follows:
In addition to the K\"ahler modulus $k_D$, there is the
K\"ahler modulus, $k_E$, of the elliptic fiber of the Calabi-Yau
manifold.  At $k_D = 0$ only a $2$-cycle collapses.  In order to
collapse the $4$-cycle one has to pass to $k_D < 0$ and go to the
point where $k_D + k_E = 0$ \KMV.  Phase $I$ is the region
with $k_D \ge 0$, and this phase connects directly to the
weakly coupled heterotic theory.  Phase $II$ is
the region for which $- k_E \le k_D \le 0$.
Common to both phases is the strong coupling singularity
where $k_D = 0$, which corresponds to an $SU(2)$  gauge
theory.   In phase $I$, one can view the low mass sector
as this gauge theory coupled to $N_f = 8$
hypermultiplets associated with the other $2$-cycles of
the del Pezzo surface.  There is a Coulomb modulus and
an effective coupling, and the latter is independent of the
former.

{}From the point of view of the five dimensional
$SU(2)$ gauge theory, phase II is the regime
where one of the hypermultiplet masses is taken to be large compared to
the expectation value of the scalar in the vector multiplet.  For
most of this paper, we will assume that this hypermultiplet mass is
infinite, leaving
an effective $SU(2)$ gauge theory with $N_f<8$ fundamental matter.  The
various $E_n$ theories then correspond to $N_f=n-1$.  There is an extra
mass parameter that appears because of the existence
of a soliton with $SU(2)$ gauge
charges for the five dimensional theory and this mass parameter combines
with the bare mass parameters for the fundamental matter to fill out the
full $E_n$ global symmetry.

For the Calabi-Yau compactification of the IIA theory, the
five dimensional gauge theories are compactified
to four dimensions on a circle of radius $R_5$.
Since these are $N=2$ theories with a one-dimensional coulomb branch,
there should exist Seiberg-Witten
elliptic curves whose modulus is the effective coupling
for the theory.  In this paper we will construct the explicit curves
for all values of $N_f<8$.  These curves have an $R_5$ dependence
which generalizes the results in \JMDN.  Varying $R_5$ from small radius to
large radius interpolates between the four dimensional superconformal
theories in \JMDN\ and the five dimensional $SU(2)$ gauge theories discussed
in \NSeib.

In the underlying string theory, phase $II$ is rather more exotic
than phase $I$, and was the focus of
\LMW.  
In particular,
it is the phase in which one can directly access the
perturbative non-critical string.  One can view $k_D$ and
$k_E$ as different combinations of the tension of
the string and the compactification radii.  In particular,
if a membrane wraps the elliptic fiber with degree $d_E$,
and the canonical divisor with degree $d_D$, then these
integers represent the winding number and momentum,
respectively, of the compactification of the non-critical string
on a circle.    The remaining
K\"ahler moduli of the del Pezzo $2$-cycles can then be
interpreted  as Wilson line parameters on the circle
of compactification.  For  $d_E > 0$ one finds that the only BPS
membranes have $d_E \ge d_D$.  The states with $d_E =
d_D = d$ are those that become massless when the $4$-cycle
collapses when $k_D + k_E = 0$.  This is an
infinite tower of states, indexed by $d$, and they
should represent a fundamental ``electric'' representation of
the non-critical string.  It is this infinite tower that
we will study extensively here.

To study these very stringy BPS states, one can use the
mirror map as in \KMV, and obtain the degeneracies
of the states.   So far only the lowest level of the
excitations ($d_E =1$) has been adequately
understood \KMV.   The problem is that it is very
hard to extract detailed information about
the $E_n$ structure from a mere count of the
number of BPS states at a given level.
One of the purposes of this paper is to describe
and utilize a technique that provides
much more precise and detailed information
about the $E_n$ structure of the spectrum.  We will
also extract some apparently universal data about
the degeneracies. One could,
in priniciple accomplish this by introducing the K\"ahler moduli
corresponding to the Wilson lines,
and then performing the mirror map.  However, we
will describe a much simpler approach
that involves passing first to a form of
consistent truncation of the type II compactification
down to the essential sector of the non-critical string \LMW.

The basic idea  of \LMW\ was to isolate the non-critical string as
a closed monodromy subsector of the type II compactification. That is,
one takes the view that one can consistently truncate a theory to a
subsector if one isolates a set of BPS states and moduli so that
monodromies over that moduli space have a closed
representation on the selected BPS states.
In \LMW\ it was proposed that one could exactly capture and
model the non-critical $E_n$ strings
by using some very particular non-compact
Calabi-Yau manifolds to ``compactify'' the IIB theory.
The approach is directly parallel to the manner in which
IIB compactification on ALE fibrations over $\IP^1$ captures
the exact quantum effective action of $N=2$ supersymmetric
gauge theories (completely decoupled from the rest of
the original type II
string theory) \KLMVW.     The claim is that by using
the proper non-compact manifolds, one can study {\it  in isolation}
the non-critical strings decoupled from the ``superfluous''
excitations of the original and larger string theory in which
the non-critical string appeared.   Such a IIB compactification
gives rise to an ``effective action'' for the non-critical string.
The natural expectation for such an effective action is that
it will describe the Coulomb branch of the gauge theory
and indeed, such field theory actions were constructed
in \refs{\JMDN,\GMS, \NSeib}.  However, getting the effective action
via a IIB compactification leads to a much deeper stringy insight
as in \KLMVW:  The BPS states appear as $3$-branes wrapping $3$-cycles,
but one can ``see'' the string by first wrapping the $3$-branes
over $2$-cycles.  The effective action is constructed
from the period  integrals of the holomorphic $(3,0)$-form on
$3$-cycles.  For the non-compact Calabi-Yau manifolds
of \LMW, these integrals can be reduced to integrals of a
meromorphic $(1,0)$-form, or generalized Seiberg-Witten
differential, on a torus.  This torus can then be thought
of as a compactifying space of the string, and the Seiberg-Witten
differential represents a local string tension.
Apart from satisfying some of the basic properties
of their their gauge theory counterparts, this formulation
of the  quantum effective actions of the non-critical strings
also has some fundamentally new features:  For example,
the differential (local string tension) vanishes indentically
at one value of the modulus (the tensionless string point) and
the asymptotic expansion of these actions yields a generating
function for the counting of BPS states.

It is important to remember that this formulation of
the non-critical string is not strictly derived from
other formulations: it was proposed for rather
general reasons, and this proposal has been checked against
the results from the mirror map.
One of the purposes of this paper is to perform extensive
further verification of this proposal by including
Wilson lines for the compactified non-critical string.
In the  effective action these parameters become masses
for the hypermultiplets for the $SU(2)$ gauge theory.
We are thus able use the techniques introduced in \SW\ and
developed in \JMDN.  We find that these masses enter the counting
of BPS states in precisely the same manner as multiple K\"ahler moduli
appear when one uses mirror symmetry to count rational curves.
Thus at one level we have found an extremely efficient method
of computing the mirror map (with up to nine parameters)
by working on a torus.
This not only enables the counting of curves in terms of
$E_n$ characters, but also enables us to study the flow
of the effective action as one moves from $E_n$ to $E_{n-1}$.
The fact that the asymptotic forms of effective actions
flows in exactly the proper manner provides strong confirmation
of our results and of the proposal of \LMW.

In section 2 we briefly review the results of \LMW, focussing
on how the effective action of the non-critical string
is computed and reduced to period integrals on a torus.
We also generalize this effective action
to include two or three mass parameters.
In section 3 we compute the instanton expansion from the
effective action, and show how this expansion is refined into
characters of $E_n$.  In section 4 we focus on
the $E_8$ theory, and reverse the previous philosophy
by using the $E_8$ structure of the instanton expansion
to determine the exact form of the torus for the
complete set of eight mass parameters.  In section 5
we make a much more extensive study of the characters
that appear in the instanton expansion, determining
degeneracies of $E_n$ Weyl orbits up to curves of degree
$5$ for  $n=8$, degree $6$ for $n=6,7$ and degree $10$
for $n=5$. We also show how one can flow from the
$E_8$ theory down to any $E_n$, and that the instanton
expansion behaves appropriately under this flow.  We
then extract some universal ($E_n$ independent) information
about degeneracies from this data.  In section 6 we show
how our methods should generalize to yield an effective
action that includes another modulus, and this should
enable the computation of the full set of excitations
of the non-critical string given in \KMV.  Section
7 contains some brief concluding remarks.  There are
two appendices:  The first contains details
of how the tori and Seiberg-Witten differentials for
the massive theories were computed.  The second
contains the explicit formulae for the tori for $E_n$,
$0 < n \le 8$.

\newsec{The Effective action of the Non-Critical String}

\subsec{The massless theory}

To compute the effective action of the non-critical, $E_n$
string it was argued in \LMW\ that one should compute the
classical pre-potential of a IIB compactification on
particular non-compact Calabi-Yau $3$-folds that depends
upon the choice of $E_n$.  Here we will restrict our
attention to $E_8$, but the calculation for other
$E_n$ proceeds similarly.

The appropriate $3$-fold is given by the following
polynomial in weighted projective space:
\eqn\Eeight{w^2 ~=~ z_1^3 ~+~ z_2^6 ~+~ z_3^6  ~-~
{1 \over z_4^6} ~-~ \psi\, w  z_1 z_2 z_3 z_4\ .}
As was described in \LMW, the Seiberg-Witten differential
for the underlying $SU(2)$ gauge theory can then be computed
by (partially) integrating the holomorphic $3$-form,
$\Omega_{(3)}$, of this
surface over suitably chosen $3$-cycles.  One can then
express the result as
\eqn\indefint{\eqalign{\lsw ~=~ &\bigg(~\int^\psi {d\zeta \over
\sqrt{1 + x^3 + \coeff{1}{4} \zeta^2 x^2}} ~\bigg) ~dx
 \cr  ~=~ & {1 \over 2}~
\log~\left[ {\sqrt{1 + x^3 + \coeff{1}{4}  \psi^2 x^2} ~+~
\coeff{1}{2} \psi x \over \sqrt{1 + x^3 + \coeff{1}{4} \psi^2
x^2} ~-~ \coeff{1}{2} \psi x } \right]~{d x \over x}  \ .}}
One should interpret $\lsw$ as a differential on the
curve $ y^2 = 1 + x^3 + \coeff{1}{4}  \psi^2 x^2$.

One can easily make a direct connection between this and the
approach of \JMDN, which is based on the more standard description
of the $E_8$ torus.  Make the change of variables:
\eqn\rescale{x ~\to~ 2^8 ~\psi^{-10}~x \ , \quad  y ~\to~ 2^{12}~
\psi^{-15}~y  \ , \quad u ~\equiv~ - \coeff{1}{32}~\psi^6 \ .}
One then obtains the curve
\eqn\canoncurve{y^2 ~=~ x^3 ~-~ 2~u^5 ~+~  u^2 x^2 \ ,}
while the differential that is being integrated in
\indefint\ becomes
\eqn\twoform{\Omega_{(2)} ~=~ {dx ~du \over \sqrt{x^3 - 2 u^5 ~+~
u^2 x^2}} \ .}
This is very close to the starting point of \JMDN\ where
the Seiberg-Witten differential is constructed by
writing such a $2$-form as the exterior derivative of
a $1$-form.  Here we see that the corresponding $2$-form
naturally appears in the partial integration of the
Calabi-Yau holomorphic $3$-form.
There are however some fundamental differences:  the curve
\canoncurve\ contains an extra term compared to that
of \JMDN. Upon shifting $x \to x - {1 \over 3} u^2$
one obtains a $u^6$ term, that
is more characteristic of the elliptic singularity, rather
than of the $E_8$ singularity.  This, combined with
the fixed normalization of $\Omega_{(2)}$, also leads
to a Seiberg-Witten differential that has irremovable
logarithmic branch cuts \indefint\ \LMW.

The Calabi-Yau manifold and the torus described above
only depend upon one complex modulus. As was described in
\LMW, this  modulus corresponds to the K\"ahler
modulus $t_S = i(k_D + k_E)$ of the IIA compactification.
The simplest closed sub-monodromy problem is the
truncation to the study of this single modulus in phase $II$.
It is also precisely this modulus that one needs to
characterize and count the fundamental massless
tower of electric states with $d_E = d_D = d$.  We
will discuss in section 6 how to restore the second
modulus to the foregoing model.

This brings us to an important technical point:
We have specialized to a one parameter closed
sub-monodromy problem based upon a single
complex structure modulus, $\psi$.  As was described in
\LMW, such a single parameter truncation only
exists in phase $II$.  On the other hand we are ultimately
going  to look at the large complex structure limit and this
corresponds taking the string tension to
infinity.  If one is in phase $II$ and one takes this
limit, then one necessarily crosses the
boundary into phase $I$, {\it unless} one is at a point
in moduli space where this boundary has been shifted
infinitely far away.  This means that the foregoing Calabi-Yau
manifold and torus must describe a rather singular limit of the
IIA compactification:  one in which
$k_E$ has been shifted off to infinity.  In terms
of the toroidal compactification of the non-critical string,
the ratio $R_5/R_6$ has been taken to infinity in such a
manner that  $\phi R_5 R_6$ remains finite, where
$\phi$ is the non-critical string tension
\foot{In terms of the torus of \GMS, we have specialized
to the point in moduli space with $\sigma = i \infty$.}.
One can view this
limit as degenerating the six-dimensional
theory to five dimensions, and then compactifying the
theory to four dimensions on a circle of radius $R_5$.
Thus there is only one scale left in the theory, namely
$1/R_5$.

It is important to highlight the unusual but crucial
form of \indefint.  As was emphasized in \LMW, for
$\psi = 0$, the differential $\lsw$
vanishes identically over the entire Riemann surface.
This is essential since the
BPS states become massless when the $4$-cycle collapses.
Moreover, if one were to obtain the Seiberg-Witten
differential by integrating the holomorphic differentials, then
apart from normalization issues, the boundary condition that $\lsw$
must vanish at $\psi = 0$ provides a constant of integration
that is crucial to the proper instanton expansion at $\psi = \infty$.
The unusual feature in \indefint\ is the presence of the logarithm,
and the logarithmic branch cuts, which imply that it is
multi-valued on the torus.  Given the geodesic
description of BPS states \KLMVW\ the multi-valuedness, at first
sight, seems extremely surprising.  However, one should recall
the rather singular limit that one has implicitly taken, and
use the fact that the only mass scale in the problem is $1/R_5$.
This scale must multiply \indefint.  The multi-valuedness of
the logarithm then implies that on the $N^{\rm th}$ sheet
the differential, $\lsw$, has
a simple pole of residue $2 \pi i N/R_5$.  Following the
rules of \SW\ this means that there must be an infinite
tower of hypermultiplet states of masses $2 \pi i N/R_5$.
These are simply the Kaluza-Klein modes of the string on the
compactified $R_5$.  If the six-dimensional theory were compactified
on a non-degenerate torus then the Seiberg-Witten differential must
involve the inverse of a doubly periodic function, that is
an inverse elliptic function whose $\tau$-parameter is that
of the torus upon which the compactification is made
(see section 6).

The logarithmic branch cuts also play an important role in the
Seiberg-Witten differential for the model with Wilson lines:
The differential must have residues that are linear in the
masses, or Wilson line parameters, $m_i$, while the parametrization
of the relevant algebraic surface must respect the periodicity
of the Wilson line parameter space, {\it i.e.} $m_i \to
m_i + 2 \pi$.  This means that the coefficient functions
in the Seiberg-Witten differential must involve inverse trigonometric
functions (or inverse elliptic functions for the toroidal
compactification).  This is exactly what one finds in \indefint.

\subsec{Incorporating masses: first iteration}

Following \refs{\SW,\JMDN} one builds the model with non-zero masses
by making deformations of \canoncurve, and seeking the
lines in the surface defined by $(y,x,u)$.  The Seiberg-Witten
differential is determined by finding $\lsw$ such that
$\Omega_{(2)} = d \lsw$ on this surface with the lines
excised.  There is still some ambiguity in this process,
but this is resolved by requiring that $\lsw$ has the proper
(Weyl invariant) residues.

If one introduces $p$ mass parameters, the $E_8$ symmetry is broken
to $SO(16 - 2p)$, and this means that the discriminant of the
curve must behave as $\sim u^{10-p}$ as $u \to 0$.  For three or
fewer masses, the general form of the curve is not very complicated.
Consider the limit $R_5=0$.  The curves in this limit were constructed
in \JMDN\ and will henceforth be referred to as the
{\it polynomial} curves because of their polynomial dependence
on the masses.  Correspondingly, we will refer to the
curves that we are about to construct as the {\it trigonometric}
curves.  One can make an Ansatz that all coefficients of $u$ and $x$
that are absent in the polynomial case remain so
for the trigonometric curves, with the exception of the coefficient of
the $u^2x^2$ term, which
is set to 1.  The non-zero terms need to be modified, but this is done
such that a Seiberg-Witten differential can be constructed whose
residues are linear in the mass parameters $m_i$.
Details of these constructions for up to two non-zero masses,
along with the corresponding
Seiberg-Witten differential are given in appendix A.

The two mass curve has a particularly simple form and is given by
\eqn\twomass{
y^2=x^3+u^2x^2-2u(u^2+\sin^2(m_+)x)(u^2+\sin^2(m_-) x), }
where $m_\pm=(m_1\pm m_2)/2$.  This can be compared with the
$E_8$ polynomial curve
\eqn\twomassconf{
y^2=x^3 -2u(u^2+ m^2_+x)(u^2+ m^2_-x),}
For   three non-zero masses we find that the curve still has the
simple form
\eqn\threemass{ y^2 ~=~ x^3 ~+~ u^2 x^2  ~-~ u ~\Big(~2 u^4 ~+~
T_2~u^2 ~x ~+~ 2 \wT_4~x^2 \Big)  ~-~ T_6~u^4 \ ,}
where
\eqn\Tparams{\eqalign{T_2 ~&\equiv~ \sum_{i = 1}^4 ~\sin^2 (p_i) \ ,
\qquad \wT_4 ~\equiv~ \prod_{i = 1}^4 ~\sin (p_i) \ ,
\qquad T_6 ~\equiv~ \prod_{i = 1}^3 ~\sin^2 (m_i) \ ; \cr
p_1 ~&\equiv~ \coeff{1}{2}( m_1 - m_2 - m_3) \ , \qquad
p_2 ~\equiv~ \coeff{1}{2}( - m_1 + m_2 - m_3) \ , \cr
p_3 ~&\equiv~ \coeff{1}{2}(- m_1 - m_2 + m_3) \ , \qquad
p_4 ~\equiv~ \coeff{1}{2}( m_1 + m_2 + m_3) \ .}}
The Seiberg-Witten differential for the two-mass curve
can be found in Appendix A.  Its form is rather complicated,
and indeed we will not need it directly -- we will only
need the asymptotic form of \indefint.

\newsec{The instanton expansion}

One of the interesting things about the effective
action defined in \LMW\ is its behaviour at large $u$.
Following \SW\ one defines
\eqn\aandaD{\eqalign{
\phi (u)\ &=\   \int_{\gamma_a} \lsw \ =\
\int \Big( \int_{\gamma_a}\omega \Big) ~du ~+~ \delta \cr
\phi_D(u)\ &=\ \int_{\gamma_b}\lsw \ =\
\int \Big(\int_{\gamma_b}\omega\Big) ~du ~+~ \delta_D \ ,
}}
where $\omega = dx/y$ is the holomorphic differential
on the torus \canoncurve, and $\delta,\delta_D$ are
integration constants.
The constants of integration are crucial to the
asymptotic expansion at infinity, and can be determined
by a careful asymptotic expansion of the period integrals
of \indefint \foot{The key observation to getting the
asymptotic expansion correct is that the log branch cuts
must run through the square-root branch cuts thereby
connecting the log branch points on different
$y$-sheets.  The curve $\gamma_a$ must not cross the
log cut.  All this is required to have the proper
$\psi \to 0$ limit, and in the $\psi \to \infty$ limit
it gives $\phi$ a $ \log(\psi)$ divergence.}, or equivalently
by analytically continuing and imposing the requirement
that $\phi$ and $\phi_D$ vanish at $\psi = 0$.

In \LMW\ it was shown that the Yukawa coupling,
$C_{\phi \phi \phi} = \partial_\phi^3 {\cal F}$,
was exactly the one obtained in \KMV:
\eqn\Einstant{-1 ~+~ 252~{1^3 q^1 \over 1 - q^1} ~-~
9252~{2^3 q^2 \over 1 - q^2} ~+~ 848628~{3^3 q^3 \over 1 - q^3} ~-~
114265008~{4^3 q^4 \over 1 - q^4} ~+~ \dots \ . }
The corresponding pre-potential, ${\cal F}$, was also shown to be:
\eqn\prepot{{\cal F} ~=~ ={1 \over 6} \phi^3 ~+~ {1 \over 4}
\phi^2 ~-~ {5 \over 12 }\phi ~+~{1 \over 4 \pi^2}\sum_{k=1}^\infty\,
n_k^r\, Li_3(e^{2\pi i k\, \phi })\ , }
where the instanton coefficients $n_k^r=\{252,-9252,..\}$.
The fact that the torus \canoncurve\ and the differential
\indefint\ replicate the counting of BPS states of
the non-critical string provided confirmation
that the foregoing does indeed provide a model of
the non-critical string.

We now describe in a little more detail how to compute
the instanton expansion from the torus, but this time
we include the two or three non-zero mass parameters
by using the torus \twomass\ or \threemass.
The first step is to recast the torus in canonical form:
\eqn\Wtorus{\eqalign{ y^2 ~=~ & 4 x^3 ~-~ g_2(\sigma)~x ~-~
g_3(\sigma) \ , \cr
g_2(\sigma) ~=~ & 60 ~\omega_2^{-4}~G_4(\sigma) \ , \quad
g_3(\sigma) ~=~ 140 ~\omega_2^{-6}~G_6(\sigma) \ , \cr
G_{2k} (\sigma) ~\equiv~& {2(2 \pi i)^{2k} \over(2k-1)!}~\Big[
\sigma_{2k-1} (n) ~q^n ~\Big]  \ ,}}
where $G_{2k}$ are the canonically normalized Eisenstein
functions, $\omega_2$ is one of the torus periods and
$q = e^{2 \pi i \sigma}$.  The other torus period is
thus $\omega_1 = \sigma \omega_2$.  This gives one expressions
for $g_2$ and $g_3$ in terms of $u$ and the $m_i$, and
substituting these into:
\eqn\jfunction{\eqalign{j(\sigma) ~=~ & {1728~g_2^3 \over g_2^3
- 27 g_3^2} \cr ~=~ & {1 \over q~\prod_{n=1}^\infty ~
(1 - q^n)^{24}}~ \Big[~1 ~+~ 240~\sum_{n = 1}^\infty ~\sigma_{2k-1}
(n) ~ q^n ~\Big]^3 \ ,}}
yields a relation between $\sigma$, $u$ and the $m_i$.  One can
then expand this in a series for large $u$, or $\sigma \to i
\infty$, and invert it to get an expansion for $\sigma$ in terms
of $u$ and the $m_i$.  Using this in $G_4$ in \Wtorus\
yields an expansion for $\omega_2$, and
hence  $\omega_1$ in terms of $u$ and the $m_i$.  To get $\phi(u)$
of \aandaD\ one integrates $\omega_2$ with respect
to $\psi$, or $u$, (using the constant of integration of
\LMW).  Inverting this one can finally determine $\sigma$,
and hence $C_{\phi \phi \phi}$ as a function of $\phi$ and
$m_i$.

The result is a series like \Einstant\ but with the
integer coefficients replaced by polynomials in
$\sin(m_i)$ or $\sin(p_i)$.  These polynomials are
then easily recognized in terms of characters of $E_8$,
or more precisely of characters of the $SO(2k)$ subgroup
of $E_8$ defined by the non-zero $m_i$, $i = 1,\dots,k$.
This information is
more than adequate to reconstruct the complete
$E_8$ characters of the first few terms of the expansion
-- indeed it is a highly overdetermined system providing
quite a number of consistency checks.

The first few terms become
\eqn\instantchar{\eqalign{
-1 ~+~ & 1^3~\big[12~\chi_{0,1}(q; m_i)  + \chi_{2,240}
(q; m_i)~\big] \cr  ~-~ & 2^3~ \big[ ~132~\chi_{0,1}(q^2; m_i)
+  20~ \chi_{2,240}(q^2; m_i)  +  2~\chi_{4,2160}(q^2; m_i) ~
\big] \cr
{}~+~ &3^3~\big[ ~4068~\chi_{0,1} (q^3; m_i) +
927~\chi_{2,240}(q^3;  m_i)  +  180~\chi_{4,2160}(q^3; m_i) + \cr
& \qquad \qquad \qquad 27~ \chi_{6,6720}(q^3; m_i)  +
3~\chi_{8,17280} (q^3; m_i) ~ \big]
 ~+~  \dots \ , }}
where
\eqn\funnychar{ \chi_{p,k}(q; m_i) ~\equiv~ \sum_{\vec v \in
{\cal O}_{p,k}}~ {q ~ e^{2i \vec v \cdot \vec m} \over
1 ~-~ q ~ e^{2i \vec v \cdot \vec m}} \ .}
In this expression, $\vec v$ is summed over the set
${\cal O}_{p,k}$, consisting of vectors with
length-squared $p$, that lie in a single
Weyl orbit of order $k$ on the root lattice of $E_8$.

One of the interesting features of \instantchar\ is the form
of the terms that subtract of the multiply wrapped rational
curves of lower degree.  In \Einstant\ this subtraction
was performed by the denominators $(1 - q^n)$, whereas
in \instantchar\ these denominators have been replaced
by
$(1-q^n e^{2i \vec v \cdot \vec m})$ for a particular $\vec v$ in
a Weyl orbit.  Expanding this denominator leads to Weyl orbit
 characters that are evaluated at $\ell m_i$,
where $\ell$ is the multiplicity of the wrapping of
the fundamental rational curve.  Thus the character
parameter properly reflects the multiple wrapping.
Indeed, the form of \funnychar\ is precisely the
proper form for effective potential obtained from a mirror map
in which $\sigma$ and the $m_i$ are K\"ahler moduli.

Thus far we have only needed the curve with three
non-zero Wilson lines. To get the complete curve
we now reverse the foregoing procedure and determine
the curve by requiring that the higher terms in the
instanton expansion have the proper $E_8$ structure

\newsec{Deriving the curves from the instanton expansion}

In the last section we saw that, in principle, the Seiberg-Witten
curves can be derived by looking for the holomorphic lines.
In practice, this can be carried out for a small number of
masses, but becomes exceedingly difficult beyond three masses.

We could in principle also derive the curves by solving the linear
equations in \GMS\ and then taking the appropriate limit to reduce
everything to the five-dimensional theory.  Unfortunately, this
too cannot be easily carried out.

We propose another way to compute the curves, which takes advantage of
the instanton expansion.  This turns out to be an efficient method.
We find that the curves are much simpler than one would expect
for the full six-dimensional theory, and in fact are not much more
complicated than the curves found in \JMDN\ for the  polynomial
mass cases.

Our strategy is to compute the instanton expansion for a curve
with unknown coefficients, assuming that the curve has the correct
polynomial limit.  We then assume that the instanton expansion will lead
to an expansion in characters for the appropriate $E_n$ group. As
we saw in the previous section, the character expansion is consistent up
to two or three masses.  It will
turn out that we will only need to impose some rather simple
constraints arising from general $E_n$ character requirement.

We will do the calculation for the  $E_n$ theories with
$n \le 8$.  One could
derive each curve individually, or one could start with the
$E_8$ curve and reduce to the lower cases by taking the masses
to a certain limit.  Doing the calculation both ways
provides some useful consistency checks.

To compute the $E_8$ curve, there is a useful trick that we can employ.
Once a curve has been obtained, it
is a straightforward procedure to derive its instanton expansion.  For
the lower $E_n$, we can compute the instanton expansion directly from its
curve, or we can derive it from the $E_8$ instanton expansion
\foot{The $E_6$ del Pezzo is of particular interest since it is
equivalent to the space of cubics in $\IC \IP_3$,
so the instanton expansion is giving us information about the holomorphic
curves on this surface. }.
Since the lower $E_n$ instanton expansions can be computed directly from
the $E_8$ expansion, all coefficients in front of characters also appear
in the $E_8$ expansion (although the reverse is not true).  This means
that some (although not all) of the $E_8$ coefficients can be determined
from the lower $E_n$.  This is useful since the $E_8$ instanton computation
is much more intensive than the lower expansions.

We will also find that there is a duality in the
character expansions, which is somehow related to the $T$-duality of
the original $2$-torus.  This will be described further in
section 5.

\subsec{The curve for $E_8$}

The $E_8$ polynomial curve was derived in \JMDN\ and is given in Appendix B.
The coefficients of the curve are written in terms of $SO(16)$ invariants
$T_n$, where
$$T_{2n}=\sum_{i_1<i_2..i_n}^8  m^2_{i_1}..m^2_{i_n}$$
and the $m_i$ are some bare mass parameters.  As we saw for the curve
with a only a few non-zero masses, the curve away from the polynomial
limit has the coefficients replaced with
polynomials of trigonometric functions of the bare masses.
A convenient basis for these functions are the set of $T_{2n}$
defined by taking
\eqn\Ttrig{\eqalign{
t_8 ~=~ \prod_{i=1}^8 \sin m_i, \qquad T_{16} ~&=~ (t_8)^2 \ ,
\qquad T_{2n}~=~ G_{2n}-T_{2n+2}\qquad 1<n<8 \ , \cr \quad
{\rm where}
\quad G_{2n}~&=~ \sum_{i_1<..i_n}\sin^2m_{i_1}...\sin^2m_{i_n}\ .
}}
One then finds
$$T_2 ~=~ 2-2\prod_{i=1}^8  \cos m_i  \ .$$
We also define the parameter
\eqn\Ttf{\wT_4={1\over4}T_2^2-T_4=T_2-G_2 \ .}

The instanton expansion is computed as in the previous section, but now
instead of showing that a curve gives a character expansion, we assume
that the character expansion exists and use this Ansatz to determine the
curve for 8 arbitrary masses.  Of course, the
character expansions for $E_8$ become large and unwieldy, even for
the smaller representations, so it is not practical to explicitly check
the instanton coefficients term by term to see if they fit into characters.

However, an important fact is that the maximal subgroup of $E_8$ is actually
$Spin(16)$.  This means that only the representations that are in the
same conjugacy class as the adjoint or one of the spinor representations
of $SO(16)$  appear in the instanton expansion.  The terms in the character
expansions have the form $exp(2i \vec m\cdot\vec\Lambda)$ where $\vec\Lambda$
is a point on the weight lattice.  The absence of vector representations
and one of the spinor reps, along with their conjugacy classes, simply means
that the instanton expansion is invariant under the $Z_2$ transformation
$m_i\to m_i +\pi/2$.  In terms of the expressions in \Ttrig, the $Z_2$
transformation is
\eqn\GTtrans{\eqalign{
G_{2n}&\to \sum_{m=0}^n\left(\matrix{&8-m \cr &8-n}\right)(-1)^m G_{2m}\cr
t_8&\to 1-T_2/2\qquad\qquad T_2\to2-2t_8.
}}

Surprisingly, insuring that the instanton expansion is invariant under
this transformation is sufficient to determine the complete $E_8$ curve.
Furthermore, not many extra terms appear away from the polynomial
limit.  The extra piece that should be added to the conformal curve in
the appendix is
\eqn\additional{\eqalign{
&x^2(u^2+2t_8\widetilde T_4)+x\Bigl(
2T_2t_8 u^2+(2 t_8 T_8+t_8 \widetilde T_4^2/2\cr
&+4 t_8^2-T_{12} \widetilde T_4)u
+4 t_8^2 T_6 - 2 t_8 T_{10} \widetilde T_4 +
 4 t_8^2 T_2 \widetilde T_4 + T_{14} \widetilde T_4^2\Bigr)\cr
&
-8T_8u^4-(4 t_8 T_6+8 T_{14})u^3-
(4 T_{14} T_6+8 t_8 T_{12}-2 t_8 \widetilde T_4 T_8+8 t_8^2 \widetilde T_4
-t_8^2 T_2^2)u^2\cr
&+
\bigl(4 t_8^3 T_2-  t_8^2 (8 T_{10} - 2 T_2 T_8
 + 2 T_6 \widetilde T_4 - (T_2 \widetilde T_4^2)/2) -t_8 (4 T_{12} T_6 +
4 T_{14} \widetilde T_4 + T_{12} T_2 \widetilde T_4 +
  T_{10} \widetilde T_4^2)\cr
&+ 2 T_{14} T_8 \widetilde T_4 \bigr)u\cr
&+4 t_8^4 +
  t_8^3 (4 T_2 T_6 - 4 T_8 + 2 T_2^2 \widetilde T_4 - \widetilde T_4^2) -
  t_8^2 (4 T_{10} T_6 - T_8^2 + 2 T_{12} \widetilde T_4
+ 2 T_{10} T_2 \widetilde T_4\cr
& -
     (T_8 \widetilde T_4^2)/2 - \widetilde T_4^4/16) +
  t_8 (T_{12} T_8 \widetilde T_4 + T_{14} T_2 \widetilde T_4^2
+ (T_{12} \widetilde T_4^3)/4) + (T_{12}^2 \widetilde T_4^2)/4 -
T_{10} T_{14} \widetilde T_4^2}}
In fact this term reduces to $x^2u^2$ if at least three of the masses
are zero.
In order to fully determine \additional\ it was necessary to compute the
fifth instanton in the expansion.  We will discuss instanton expansions
in more detail in the next section.
In the conformal limit, the variables have dimensions $[x]=10$, $[u]=6$
and $[T_n]=n$, so that all terms in (B.1) have dimension 30.  Using
these conformal dimensions, we see that all terms in \additional\ have
dimension 32.  This of course does not mean that all possible dimension
32 terms appear, in fact the majority of them do not.

\subsec{The curves for the other $E_n$}

With the full $E_8$ curve one can derive the curves for the smaller $E_n$.
To compute the $E_7$ curve, take $m_7=i\Lambda+\mu$ and
$m_8=-i\Lambda +\mu$ and take the limit $\Lambda\to\infty$
(which corresponds to a large mass for
the five dimensional gauge theory).
In this limit $\sin m_7\approx ie^{\Lambda-i\mu}/2$ and $\sin m_8\approx
-ie^{\Lambda+i\mu}/2$.  The $T_n$ parameters scale as
\eqn\Tres{\eqalign{
T_2&=(e^{2\Lambda}/4)(T_{2,6}-2)
\qquad\qquad\wT_4=(e^{2\Lambda}/4)(T_{2,6}-4\sin^2\mu)\cr
T_6&=(e^{4\Lambda}/16)(T_{2,6}-T_{2,6}^2/4)
\qquad\qquad t_8=(e^{2\Lambda}/4)t_6\cr
T_{2n}&=(e^{4\Lambda}/16)T_{2n-4,6}\qquad 4\le n\le7,
}}
where $T_{n,6}$ are the $T_n$ variables for the remaining six masses.
 Rescaling the $u$ and $x$ variables
\eqn\EVIIres{u\to {1\over4} e^{2\Lambda}u\qquad\qquad
x\to {1\over16} e^{4\Lambda}(x+t_6T_2)}
and keeping only the leading terms in $e^\Lambda$ one finds the $E_7$
curve given in  Appendix B.

The $E_6$ curve can be derived by scaling three masses, with
$m_6=i\Lambda_1+2\lambda/3$, $m_7=i\Lambda_2+2\lambda/3$, and
$m_8=-i(\Lambda_1+\Lambda_2)+2\lambda/3$.
Taking the limit
$\Lambda_i\to\infty$ the $T_n$ scale as
\eqn\TresVI{\eqalign{
T_2&=(e^{2(\Lambda_1+\Lambda_2)}/8)(T_{2,5}-2)
\qquad\qquad\wT_4=(e^{2(\Lambda_1+\Lambda_2)}/8)
(T_{2,5}-2+2e^{-2i\lambda})\cr
T_6&=-(e^{4(\Lambda_1+\Lambda_2)}/64)(1-T_{2,5}+T_{2,5}^2/4)
\qquad\qquad T_8=(e^{4(\Lambda_1+\Lambda_2)}/64)(-T_{2,5}+T_{2,5}^2/4)\cr
t_8&=i(e^{2(\Lambda_1+\Lambda_2)}/8)t_5\qquad\qquad
T_{2n}=-(e^{4(\Lambda_1+\Lambda_2)}/64)T_{2n-6,6}\qquad 5\le n\le7
}}
After rescaling $u$ and $x$ as
\eqn\EVIres{
u\to ie^{2(\Lambda_1+\Lambda_2)}e^{-i\lambda}u/8
\ , \qquad x\to-e^{4(\Lambda_1+\Lambda_2)}\left(e^{-2i\lambda}x+
{i\over2}T_2ue^{-i\lambda}+2it_5-iT_2t_5\right)/64,}
 the $E_8$ curve
reduces to the $E_6$ curve in the appendix.  Note that even for the massless
$E_6$ case, there are imaginary coefficients for the curve.  One consequence
of this is that the $E_6$ character expansion will be not be symmetric under
complex conjugation.

The smaller $E_n$ can be derived in a similar fashion.  The masses satisfy
\eqn\massres{m_i=i\Lambda_i +{2\over9-n}\lambda\qquad n\le i<8,\qquad\qquad
m_8=-i\sum_{i=n}^7\Lambda_i+{2\over9-n} \lambda.}
The $u$ and $x$ variables are then scaled as
\eqn\xures{
u\to \left(i\over 2\right)^{9-n}e^{2i\sum \Lambda_i}e^{-i\lambda}u\qquad\qquad
x\to \left(i\over 2\right)^{18-2n}e^{4i\sum \Lambda_i}e^{-2i\lambda}x}
It is also convenient to shift the $x$ variable in order to have
a more compact expression.  The particular shift depends on which $E_n$
theory is being considered.  The curves along with the scaling
details for these smaller $E_n$ are given in the appendix.

\newsec{Character Expansions and Holomorphic Curves}

\subsec{Rational curves in $B_n$}

The contributions of the characters of the form \funnychar\
to the instanton expansion  are given in tables
1, 2 and 3 for the $E_n$ groups.   We have expressed the characters in
terms of the Weyl orbits instead of the $E_n$ representations.  As
was shown in \claude, this is a natural way to classify the holomorphic
curves on the various del Pezzo surfaces.

The $B_n$ del Pezzo surfaces are constructed by blowing up $n$ points
on $\IC \IP_2$.  The anti-canonical divisor is given as
\eqn\acd{{\cal K}=3l-\sum_{i=1}^n e_i}
where $l$ is the anti-canonical divisor on $\IC \IP_2$, in other words,
it is a generic line,  and the $e_i$ are the $n$ exceptional divisors of
the blow-up points.  The intersection matrix is generated by
$l^2=1$, $e_i^2=-1$ and $e_i\cdot e_j=0$ if $i\ne j$.
A curve in the homology class $a_0l-\sum a_ie_i$ then has degree
\eqn\deg{
d=a\cdot\mu=3a_0-\sum a_i,\qquad\qquad \mu=(3,1,1,1,1,1,1,1,1).}
The arithmetic genus of this curve is given by
\eqn\ag{
g_a={1\over2}(a_0-1)(a_0-2)-{1\over2}\sum a_i(a_i-1),}
which counts the number of double points on $\IC \IP_2$ that are not on
the blow-up points.

The holomorphic curves can then be grouped into $U(n)$ Weyl orbits and
for $n=6,7,8$, these multiplets can be further combined into $E_6$, $E_7$
and $E_8$ multiplets.  The weight length squared for a given curve is
\eqn\wlsq{
L^2=-a_0^2+\sum_i a_i^2 +{d^2\over 9-n}}
or in terms of the degree and the arithmetic genus is
\eqn\wdag{L^2={1\over 9-n}d^2+2(1-g_a).}
The $a_i$ are non-negative integers and $a_0$ is positive, except when
the curve is one of the exceptional divisors.
In the latter circumstance one has $a_0=a_j=0$, $j\ne i$ and
$a_i=-1$ for the $e_i$ divisor.

Finally, not all combinations of $a_0$ and $a_i$ are allowed.  Obviously,
we cannot have a curve with arithmetic genus less than zero.  We also
cannot have curves where $a_0<a_i+a_j$ for any $i$ and $j$.  Otherwise,
it would be possible to have a line intersecting a curve of degree $a_0$
in $\IC \IP_2$ more than $a_0$ times, which violates Bezout's theorem.
Other constraints are
\eqn\consts{\eqalign{
2a_0&\ge a_1+a_2+a_3+a_4+a_5,\cr
3a_0&\ge 2a_1+a_2+a_3+a_4+a_5+a_6+a_7\cr
4a_0&\ge 2a_1+2a_2+2a_3+a_4+a_5+a_6+a_7+a_8\cr
5a_0&\ge 2a_1+2a_2+2a_3+2a_4+2a_5+2a_6+a_7+a_8\cr
6a_0&\ge 3a_1+2a_2+2a_3+2a_4+2a_5+2a_6+2a_7+2a_8.
}}
One consequence of these constraints is that $a_i\le d$ for $B_6$ and
$B_7$ and $a_i\le 2d$, $|a_i-a_j|\le d$ for $B_8$.

\goodbreak
{\vbox{\ninepoint{
$$
\vbox{\offinterlineskip\tabskip=0pt
\halign{\strut\vrule#
&\hfil~$#$
&\vrule#&~
#&\hfil~$#$
&\vrule#&~
\hfil ~$#$~
&\hfil ~$#$~
&\hfil $#$~
&\hfil $#$~
&\hfil $#$~
&\hfil $#$~
&\vrule#
\cr
\noalign{\hrule}
&  && & &&d&  1  &  2  &  3 & 4 & 5 & \cr
\noalign{\hrule}
&L^2 && &{\rm dim }&& &   &     &      &       &        &         \cr
& 0 && &1 && & 12 &  -132 & 4068     & - 224688 & 17720400 &   \cr
& 2 && &240 && & 1 & - 20  & 927 & -66912 & 6381850 &    \cr
& 4 && & 2160 && &   & -2  &180 &  -18496 &  2207400&  \cr
& 6 && &6720 && &   &     & 27     &  -4656  &   729000 & \cr
& 8 && &17280 && &   &     &3 & -1056 &    228890  &  \cr
& 8 && &240 && &   &     & & -976 &  226100  &  \cr
& 10 && &30240 && &   &     & & -200 &  67325  &  \cr
& 12 && &60480 && &   &     & & -32 &  18540 &  \cr
& 14 && &69120 && &   &     & & -4 &  4725 &  \cr
& 14 && &13440 && &   &     & &  &  4325  &  \cr
& 16 && &138240 && &   &     & &  &  1025 &  \cr
& 16 && &2160 && &   &     & &  & 1100  &  \cr
& 18 && &181440 && &   &     & &  &  205 &  \cr
& 18 && &240 && &   &     & &  &  &  \cr
&20 && &241920 && &   &     & &  & 35 &  \cr
& 20 && &30240 && &   &     & &  &  &  \cr
& 22 && &181440 && &   &     & &  &  &  \cr
& 22 && &138240 && &   &     & &  & 5 &  \cr
\noalign{\hrule}
& && &{\rm Total} && &252   & -9252    & 848628 & -114265008  & 18958064400
 &  \cr
\noalign{\hrule}}
\hrule}$$
\vskip-7pt
\noindent
{\bf Table 1}: Coefficients for $E_8$ Weyl orbits in instanton expansion.
The bottom line is the coefficient when all $m_i=0$.}
\vskip7pt}}

\goodbreak
{\vbox{\ninepoint{
$$
\vbox{\offinterlineskip\tabskip=0pt
\halign{\strut\vrule#
&\hfil~$#$
&\vrule#&~
#&\hfil~$#$
&\vrule#&~
\hfil ~$#$~
&\hfil ~$#$~
&\hfil $#$~
&\hfil $#$~
&\hfil $#$~
&\hfil $#$~
&\hfil $#$~
&\vrule#
\cr
\noalign{\hrule}
&  && & &&d&  1  &  2  &  3 & 4 & 5 &6 & \cr
\noalign{\hrule}
&L^2 && &{\rm dim }&& &   &     &      &       &        &       &  \cr
& 0 && &1 && &  & -20 &   &-976  &  & -179028 &  \cr
& 3/2 && &56 && &1 &  &27   &  &4325  &  &    \cr
& 2 && & 126 && &   & -2  & &  -200 & &-54894 & \cr
& 7/2 && &576 && &   &     & 3     &  &1025   &  &\cr
& 4 && &756 && &   &     & &-32 &    &- 15624 & \cr
& 11/2 && &1512 && &   &     & & & 205  & & \cr
& 6 && &2016 && &   &     & & -4 &    & -4140 & \cr
& 6 && &56 && &   &     & &  &    &-3780 & \cr
& 15/2 && &4032 && &   &     & & & 35 & & \cr
& 8 && &4032 && &   &     & &  &    &-936 & \cr
& 8 && &126 && &   &     & &  &    & -1020  & \cr
&19/2 && & 4032 && &   &     & &  & 5  && \cr
& 19/2 && &1500 && &   &     & &  &   & & \cr
& 10 && &7560 && &   &     & &  &   & -198& \cr
& 12 && &10080 && &   &     & &  &   & -36& \cr
& 12 && &1312 && &   &     & &  &   & & \cr
& 14 && &4032 && &   &     & &  &   & -6& \cr
& 14 && &12096 && &   &     & &  &   & & \cr
& 14 && & 576 && &   &     & &  &   & -6& \cr
\noalign{\hrule}
& && &{\rm Total} && &56   & -272    &3240 & -58432  &1303840 & -33255216
 &  \cr
\noalign{\hrule}}
\hrule}$$
\vskip-7pt
\noindent
{\bf Table 2}: Coefficients for $E_7$ Weyl orbits in instanton expansion.
}
\vskip7pt}}

\goodbreak
{\vbox{\ninepoint{
$$
\vbox{\offinterlineskip\tabskip=0pt
\halign{\strut\vrule#
&\hfil~$#$
&\vrule#&~
#&\hfil~$#$
&\vrule#&~
\hfil ~$#$~
&\hfil ~$#$~
&\hfil $#$~
&\hfil $#$~
&\hfil $#$~
&\hfil $#$~
&\hfil $#$~
&\vrule#
\cr
\noalign{\hrule}
&  && & &&d&  1  &  2  &  3 & 4 & 5 &6 & \cr
\noalign{\hrule}
&L^2 && &{\rm dim }&& &   &     &      &       &        &       &  \cr
& 0 && &1 && &  &   & 27     &  &  & -3780 &  \cr
& 4/3 && &27 && &1 &  &   & -32 &  &  &    \cr
& 4/3 && & 27 && &   & -2  & &  &  205& & \cr
& 2 && &72 && &   &     & 3     &  &   & -936 &\cr
& 10/3 && &216 && &   &     & &-4 &    & & \cr
& 10/3 && &216 && &   &     & & & 35  & & \cr
& 4 && &270 && &   &     & &  &    & -198 & \cr
& 16/3 && &432 && &   &     & &  &    & & \cr
& 16/3 && &432 && &   &     & & & 5 & & \cr
& 6 && &720 && &   &     & &  &   & -36& \cr
& 8 && &432 && &   &     & &  &   & -6& \cr
& 8 && &72 && &   &     & &  &   & -6& \cr
\noalign{\hrule}
& && &{\rm Total} && &27   & -54    &243 & -1728  & 15255 & -153576
 &  \cr
\noalign{\hrule}}
\hrule}$$
\vskip-7pt
\noindent
{\bf Table 3}: Coefficients for $E_6$ Weyl orbits in instanton expansion.
The bottom line is the coefficient when all $m_i=0$.}
\vskip7pt}}

For instanton number $d$, one finds
that characters with weight lengths squared up to ${1\over9-n}d^2-d+2$
contribute to the pre-potential.  These maximal weight characters
correspond to the holomorphic curves of degree $d$ and
arithmetic genus 0.  The shorter lengths
correspond to curves with non-zero arithmetic genus.

\subsec{Flowing from $E_8$ to $E_n$}

The data given in Tables 1--4 was generated by working
with the individual curves for $E_n$, $5 \le n \le 8$.
Upon inspection one notices many smiliarities in the orbit
degeneracies: namely for a given degree, any number that appears
in the $E_6$ or $E_7$ table, also appears in the $E_8$ table
(although the reverse is not true).  In retrospect, this
should not have been a surpise given that we
obtained instanton expansions like \instantchar\ in
terms of functions of the form \funnychar.

One of the beautiful features of the functional
form of \funnychar\ is that one can easily use it to study
the flows down the chain of $E_n$ del Pezzo surfaces.
If one thinks of (a purely imaginary) $m_i$ as representing
the scale of a del Pezzo $2$-cycle, or as representing a mass
of a hypermultiplet, then by taking $m_i \to i \infty$ one
decouples the corresponding states from the non-critical
string, and decouples the corresponding hypermultiplets
from the field theory.  Thus the scaling
procedure given in the last section for getting the
$E_n$ curves from the $E_8$ curve should produce the
proper instanton expansions for the $E_n$ theory.
This is indeed what we find.

To be more specific, to get the  $E_n$ curves, not only were
$9-n$ of the masses given large imaginary values, $i\Lambda_i$,
but $u$ was rescaled as well.  Recall that to leading order,
$u=exp(-2\pi i\phi)$, so the rescaling in $u$ corresponds to a
shift in $\phi$, and hence a rescaling of $q$.  Indeed, for
the scaling in \xures\ one finds  $q\to e^{-2\sum\Lambda_i}q$.
The mass shifts generate a scaling in $e^{2i\vec v\cdot \vec m}
\to e^{\sum_i v_i \Lambda_i} e^{2i\vec v\cdot \vec m}$.
Thus, depending upon which of these scalings wins out,
there are three possibilities for the function
$q^d e^{2i\vec v\cdot \vec m}/(1-q^d e^{2i\vec v\cdot\vec m})$
in \funnychar: (i)  it vanishes, (ii) it is independent
of the $\Lambda_i$, or (iii) it goes to $-1$.
If the last possibility is realized then it generates
a contribution to the constant term at the front of
the instanton expansion.  To be consistent with the anomaly
computation in the five-dimensional field theory \NSeib, this
constant must change from $-1$ in the $E_8$ theory to
$(n-9)$ in the $E_n$ theory.

To find out what happens to the contributions of the various vectors,
$\vec v$, under the rescaling, it is most convenient to work in the
basis described in section (5.1).   The inner product between any two
vectors $a_1$ and $a_2$ that correspond to rational curves is
\eqn\inner{
(a_1,a_2)=-(a_1 - d_1\mu)\cdot(a_2 - d_2\mu)=d_1d_2-a_1\cdot a_2,}
where $d_1$ and $d_2$ are the degrees for $a_1$ and $a_2$ respectively.
In this basis, it is clear that the mass shift vector can be chosen to be
\eqn\massshift{
i\Lambda=\sum_{i=1}^{8-n} \Lambda_i e_{n+i}}
Hence, the inner product of a vector $a$ with $i\Lambda$ is
\eqn\awL{
(a,i\Lambda)=d\sum_i^{8-n}\Lambda_i - \sum_i^{8-n}\Lambda_i a_{n+i}}
Assuming that $\Lambda_i>0$, we see that $(a,i \Lambda)>d\sum_i\Lambda_i$
only if some of the $a_i$ are negative.  But this is possible only for
the $e_i$ divisors, with $i\ge n$.  Hence, the contribution of these
vectors to the rescaled instanton sum is $n-8$, exactly as required.
For all other $a$, the
$a_i$ components are non-negative, so the inner product is an equality
only if $a_i=0$ for $i\ge n$.   Hence, the characters from these vectors
will flow to $E_n$ characters and the coefficients in front of the
characters remain the same.  If any of these $a_i$ are positive, then
the corresponding contribution to the $E_8$ character flows to zero.

We can also flow to the $B_0$ surface, which is $\IC \IP_2$, with the
mass shift
$i\Lambda=\sum_{i=1}^8 \Lambda_i e_i$.  In the $SO(16)$ basis
$e_1={1\over2}(1,1,1,1,1,1,1,1,1)$.  In this case, one finds that the
instanton expansion only has contributions when $d=0\ {\rm mod}\ 3$.
Presumably, this instanton expansion is giving us information about
rational curves on $\IC \IP_2$.

The fact that this works so simply, and is completely
consistent with the results coming from the Calabi-Yau
compactifications \KMV, and from the field theory \NSeib,
gives even more support to the contention that the
effective action is faithfully capturing the structure of the
non-critical string.

\subsec{Reducing $E_8$ with real values of $m_i$}

We have jsut seen that by tuning $m_7$ and $m_8$ to large
imaginary values, we could flow from  the $E_8$ curve to the $E_n$ curve.
These values of the masses correspond to Wilson lines along the sixth
dimension.
By $T$ duality, we would expect a similar result for Wilson lines
along the fifth dimension.

In particular, consider what happens to the instanton expansion
when $m_7=m_8=\pi/2$, with all other $m_i=0$.  A straightforward
calculation gives for the Yukawa coupling
\eqn\VredI{
\partial^2_\phi\phi_D=-1+28{q\over1-q}-136{8q^2\over1-q^2}+
1620{27q^3\over1-q^3}-29216{64q^4\over1-q^4}+...}
This is the massless $E_7$ instanton expansion, up to a factor of
two.  If $m_7=m_8=\pi$, then we get back the original massless $E_8$
expansion.

Likewise, when $m_6=m_7=m_8=\pi/3$ and all other $m_i=0$, then the
$E_8$ characters lead to the expansion
\eqn\VredII{
\partial^2_\phi\phi_D=-1+9{q\over1-q}-18{8q^2\over1-q^2}+
81{27q^3\over1-q^3}- 5085{64q^4\over1-q^4}+...}
which is the $E_6$ expansion, up to a factor of three.

\subsec{General Structure of the Instanton Expansion}

Rational curves of degree $d$ in $B_n$ are not isolated for
$d \ge 2$:  they have moduli spaces of dimension $d-1$.  In
\refs{\MKYuM,\claude} the curve counting was stablized
by requiring that the curves pass through $d-1$ points
in general position.  If the curve has arithmetic genus
equal to zero, then according to \claude\ this imposes
an additional $d-1$ linear constraints.  This suggests
that the moduli space of curves of arithmetic genus
zero is $\IC \IP_{d-1}$.   For the
curves of small degree one can easily check this explicitly.
For instance,  consider conics going through $p$ of the
blow-up points on $\IC \IP_2$. The degree of such a curve is
$6 - p$.   A general conic on $\IC \IP_2$ has the form
\eqn\conic{
0=a_1X^2+a_2Y^2+a_3Z^2+a_4XY+a_5XZ+a_6YZ \ .}
Since conics are automatically rational, there
are no constraints on the $a_i$.  Hence the moduli space
for conics on $\IC \IP_2$ is $\IC \IP_5$.  Requiring that the conic pass
through $p$ points leads to $p$ linear constraints on the $a_i$
and reduces the moduli space to $\IC \IP_{5-p}$.

The usual expectation from using mirror symmetry to count
rational curves is that if there is a non-trivial space
of moduli, then the the ``number'' of such curves is the
Euler characteristic of the moduli space.  On a more
physical level the Euler characteristic should be thought
of as a ``net number'' after some deformation has broken
the continuous degeneracy of the space of rational curves.
At any rate, since the Euler characteristic of $\IC \IP_{d-1}$
is $d$, and the degeneracy of rational curves of arithmetic
genus zero is indeed $d (-1)^{d+1}$ (where the sign is due to
the embedding of the holomorphic curve), we seem to have some
agreement with what one expects from mirror symmetry.
However, this only ``explains'' the counting of curves
of arithmetic genus zero.

The computation of the character
expansions within the instanton expansion gives us the
ability to a take a family of curves of a given degree
and separate out curves of different arithmetic genus:
for a given degree, the length-squared of the vector
on the root lattice decreases with the arithmetic
genus.  Moreover, it is possible to have more than one Weyl
orbit of vectors of a particular length, and starting
at arithmetic genus three, these different orbits can come
with different non-zero coefficients in the instanton expansion.
As a result we should
be able to  make a finer distinction between the various
parts of the  moduli spaces that contribute to the entire moduli
space of curves, and somehow see this reflected in the
computation of the Euler characteristic.  It has
been suggested that what we are seing is the Euler
characteristic of different ``stratifications of
the moduli space'' \DMPC.

In order to try to get some control over the
large amount of data that we have gathered, we now try
to extract some universal information. As we have
seen, the longest weights at a given degree $d$ have degeneracy
$d$.  We  also note that the second longest weights also fit
into a pattern. Except for the $d=1$ case,
this coefficient appears to be $12d-d^2$.  These are the characters that
correspond to the curves with arithmetic genus $1$.

We have also found  patterns for the curves with $g_a\le 4$,
and indeed have found polynomials that fit the
Weyl orbit degeneracies. In order to
find these polynomials, it is necessary to compute the instanton expansions
at least up to order 10.  This is impractical for $E_6$ and higher
but is possible for $E_5=SO(10)$.  The $SO(10)$ case will not contain
all of the coefficients, but it contains enough to at least study
the curves for $g_a\le 4$.  Table 4 contains the $SO(10)$ coefficients
for degrees 7 through 10.  Using these numbers and the lower numbers
obtained from the $E_8$, $E_7$ and $E_6$ curves, we can construct Table
5.

As with $g_a=1$, the
the first number in each row actually violates
the polynomial rule.  Given this,
the reader might be surprised to note that we were able to derive the
second fifth order polynomial for the $g_a=4$ case with just two data points
(the first point is assumed to violate the rule).
However, we actually have more information, since we assumed that the
unknown coefficients were positive integers for $d<15$.  If we also use
the Ansatz that the polynomial contains the product of two quadratic
polynomials, then there is a unique result.

An interesting fact about the instanton expansion is that not
all Weyl orbits appear in the expansion at a given instanton number $d$,
even if other orbits of equal or greater length appear.  For instance,
for $d=3$ in the $E_8$ case, the 240 of $L^2=8$ does not appear.  As it
so happens these holomorphic curves do not exist, since they violate
the constraints in \consts.  Another interesting observation about these
holomorphic curves is that curves with arithmetic genus zero appear
at all degrees, but seem to have an upper degree limit for $g_a\ne 0$.
For instance, for $g_a=1$, there are no curves with $d>9$.  The $g_a=1$
curves have the coefficients $12d-d^2$, hence this number never changes
sign.  We expect a similar result for higher values of $g_a$.

\goodbreak
{\vbox{\ninepoint{
$$
\vbox{\offinterlineskip\tabskip=0pt
\halign{\strut\vrule#
&\hfil~$#$
&\vrule#&~
#&\hfil~$#$
&\vrule#&~
\hfil ~$#$~
&\hfil ~$#$~
&\hfil $#$~
&\hfil $#$~
&\hfil $#$~
&\hfil $#$~
&\hfil $#$~
&\vrule#
\cr
\noalign{\hrule}
&  && & &&d&  7  &  8  &  9 & 10 & & & \cr
\noalign{\hrule}
&L^2 && &{\rm dim }&& &   &     &      &       &        &       &  \cr
& 0 && &1 && &  & -9604  &    &  &  &  &  \cr
& 5/4 && &16 && & &  &25758   & &  &  &    \cr
&1 && &10 && &   &   & & -181550 & & & \cr
&5/4 && &16 && &812   &     &      &  &   & &\cr
&2 && &40 && &   &  -2752   & & &    & & \cr
&13/4 && &80 && &   &     &7992 & &  & & \cr
&3 && &80 && &   &     & & -61700 &    &  & \cr
&13/4 && &80 && &182   &     & &  &    & & \cr
&4 && &10 && &   &-768     & & & & & \cr
&4 && &80 && &   &-672     & &  &    & & \cr
& 5 && &16 && &   &     & &-17770  &    & & \cr
& 5 && &16 && &   &     & &-20000  &   && \cr
& 5 && &80 && &   &     & &-20750  &   & & \cr
&21/4 && &160 && &   &     &2106 &  &    & & \cr
&21/4 && &160 && &35   &     & &  &    & & \cr
&6 && &240 && &   &-160     & &  &    & & \cr
&7 && &320 && &   &     & &-5700  &   & & \cr
&29/4 && &80 && &   &     &630 &  &    & & \cr
&29/4 && &160 && &   &     &531 &  &    & & \cr
&29/4 && &80 && &7   &     & &  &    & & \cr
& 8 && &40 && &   & -32    & &  &   && \cr
&9 && &10 && &   &     & &-2250  &   & & \cr
&9 && &240 && &   &     & &-1550  &   & & \cr
&37/4 && &320 && &   &     &135 &  &    & & \cr
& 10 && &80 && &   & -8    & &  &   && \cr
&11 && &240 && &   &     & &-500  &   & & \cr
&11 && &160 && &   &     & &-400  &   & & \cr
&45/4 && &16 && &   &     &27 &  &    & & \cr
&13 && &80 && &   &     & &-110  &   & & \cr
&13 && &80 && &   &     & &-110  &   & & \cr
&53/4 && &80 && &   &     &9 &  &    & & \cr
&17 && &80 && &   &     & &-10  &   & & \cr
\noalign{\hrule}
\noalign{\hrule}}
\hrule}$$
\vskip-7pt
\noindent
{\bf Table 4}: Coefficients for $SO(10)$ Weyl orbits in the instanton
expansion.
We have only included those Weyl orbits that contribute up to $d=10$.}
\vskip7pt}}

\goodbreak
{\vbox{\ninepoint{
$$
\vbox{\offinterlineskip\tabskip=0pt
\halign{\strut\vrule#
&\hfil~$#$
&\vrule#&~
#&\hfil~$#$
&\hfil ~$#$~
&\hfil $#$~
&\hfil $#$~
&\hfil $#$~
&\hfil $#$~
&\hfil $#$~
&\hfil $#$~
&\hfil $#$~
&\hfil $#$~
&\vrule#&~
\hfil ~$#$~
&\vrule#
\cr
\noalign{\hrule}
&  &&d& 1  & 2  & 3 & 4 &5 &6 &7 &8&9&10&&{\rm Polynomial}& \cr
\noalign{\hrule}
&g_a &&  &   &     &      &       &        &       &  &&&&&&\cr
& 0 &&  & 1 &2 & 3   & 4 & 5 &6  & 7&8&9&10&&d/0!& \cr
&1 && & 12 &20   &27 &32  & 35 &36 &35& 32&27& &&(12-d)d/1!&  \cr
&2 && &   &132 &180  &200 &205 &198 &182&160&135&110&&(d^2-27d+192)d/2!& \cr
&3 && &   &     & 927     &1056 &1025 &936  &812 &672 &531&400&&
(d^2-30d+248)(15-d)d/3!&\cr
&3 && &   &    & &976&1100 &1020    &? & 768&630&500&&
(d^2-29d+240)(16-d)d/3!&\cr
&4 && &   &     & &4656 &4725  &4140 &? &2752&2106&1550&&
(d^2-31d+270)(d^2-35d+312)d/4!&\cr
&4 && &   &     & &  &4325    & 3780 &? &?&?&2250&&
(d^2-19d+198)(15-d)(20-d)d/4!&\cr
\noalign{\hrule}
\noalign{\hrule}}
\hrule}$$
\vskip-7pt
\noindent
{\bf Table 5}: Coefficients for a given degree and arithmetic genus.
The question marks indicate values that are non-zero but which we did
not determine.  The Polynomial gives the $d$ dependence for the coefficients.
For non-zero $g_a$, the first term in each row violates the polynomial
rule.
}
\vskip7pt}}

We should stress that except for the $d=1$ or $g_a=0$ curves,
the coefficients in Table 5 are
not the Euler numbers for the moduli spaces of the relevant curves.
For instance the coefficient for the $d=2,\ g_a=1$ curves is $-20$,
but the Euler number for the moduli space of these curves
is $-4$.  One can derive this as follows.  Let
$F(x,y,z)=\lambda_1 f_1+\lambda_2 f_2 + \lambda_3 f_3$ be a
pencil of cubics that intersects
seven points on $\IC \IP_2$.  To reduce this pencil to the space of
{\it rational} cubics, there must be a double point, in other words
a point where $\partial_x F=\partial_y F= \partial_z F=0$.  This has
a solution for a set $\{\lambda_1,\lambda_2,\lambda_3\}$ if
the determinant  $\Delta=|\partial_i f_j|$
is zero for some point on $\IC \IP_2$. The determimant
$\Delta$ is a sixth order polynomial
on $\IC \IP_2$, which naively is a genus 10 Riemann surface.
However, $\Delta$ has a double point at each of the original
seven points, hence the genus is 3 and the Euler number is $-4$.

Even though the instanton coefficients are not the Euler numbers,
we believe that the polynomials in Table 5 contain information about
the topological structure of the spaces of moduli of the curves.
For example, the coefficients for the terms linear in $d$
are very interesting numbers:  they appear
in the work of \refs{\MKYuM,\claude}, and for a given $g_a$
represent the number of
rational curves of degree $d$ through $d-1$ specified points
in general position in $\IC \IP^2$ or $\IP^1 \times \IP^1$.
The leading term in the polynomial
 seems to have the universal form $d^{g_a+1}/(g_a)!$
and presumably reflects some combinatorial factor.
It  is amusing to conjecture that the $d^p$ coefficients of the
polynomials are related to the Euler characteristic of the $p -1$
dimensional space of rational curves of degree $d$ passing through
$d-p$ points.

As regards the ``errors'' for $g_a =1$
in the polynomials in table 5, it is tempting to try
to associate this with the fact that we are looking
at a degenerated torus compactification of the non-critical
string.  For example,  there are actually $12$ curves with
$d=1, g_a =1$, rather than the $11$ predicted by the polynomial
formula.  These rational curves can
be found explictly and correspond to the singular fibers
on the $2$-fold in $(x,y,u)$ defined by \canoncurve.
One of these fibers is at infinity in the $u$ plane, leaving 11
at finite $u$.

While the mathematical interpretation of these degeneracies
is as yet unclear, it seems very likely that one will
ultimately be able to find the proper interpretation.
However, we feel that this is not the right way to understand
the issue.  There should be a simple physical characterization
of these degeneracies, and the mathematical interpretation
will then amount to a magical property of this partition
function of the non-critical string.

{}From the physics perspective, the most important fact about
the polynomials in Table 5. are that they are universal:  that
is, they represent degeneracies for the $E_n$ string for
any $n$.    One should recall that in terms
of the compactified string, the degree $d$ represents a
winding number and momentum state of a string on a circle.
The belief is that the $d$-wound state is, in fact, a
bound state, and so these degeneracy polynomials are fundamental,
group theory independent, properties of the bound states spectrum.

\subsec{Counting states via BPS geodesics}

Our approach to the $E_n$ string has been based upon a
IIB compactification on a
Calabi-Yau manifold in which one has integrated out
two dimensions to obtain a torus.  As mentioned earlier,
this should enable us to see the string rather explicitly,
and as in \refs{\KLMVW,\WLer, \ABSS,\JRab,\JSNW}, count BPS
states by
counting indecomposable geodesics on the torus with metric
$ds^2 = |\lsw|^2$.  The existence of BPS geodesics
can be rather subtle in that the curves of minimal length with
a given  set of winding numbers could be decomposable into
concatenations closed geodesics of other winding numbers.
In such circumstances, the corresponding BPS state will
be either marginally stable or unstable.  Thus, at
strong coupling, some purely electric states can become
unstable (like the $W$-boson in $N=2$ supersymmetric gauge
theory).   However, if one is interested in the purely electric
bound states of the $E_n$ string, one might hope that if
the string tension is high enough, then all such states would
be stable against decay into magnetically charged states.  Thus
all the fundamental electrically charged excitations of
the $E_n$ non-critical strings should be counted by looking
at all the homotopy classes of strings on the torus with
winding numbers $(1,0)$ about the $(A,B)$-cycles.
We will therefore see to what extent this approach replicates
the state counting that we have already done.

To count the geodesics properly, one must of course, keep
track of the hypermultiplet charges, and this is done
by turning on all of the mass parameters and keeping track of
the winding numbers around the simple poles of $\lsw$ whose
residues are linear combinations of the $m_i$.
We will refer to such poles as
{\it relevant \ } poles.  We will ignore both the
multi-sheeting of the torus induced by the logarithm, and the
({\it irrelevant}) poles of residue $2 \pi i N/R_5$, since as
we discussed before,
this structure is related to Kaluza-Klein momenta.  The obvious
hypothesis for the set of stable, fundamental electric BPS states
is the set of curves that pass once around the $A$-cycle of
the torus, passing between relevant poles of $\lsw$.  The number of
such states is equal to the number of relevant poles of
$\lsw$.

So far in this paper we have not needed $\lsw$ for the
massive theory in order to do the BPS state counting.  Now we
need it, and we need to exploit an ambiguity in its definition.
As discussed in \refs{\SW,\JMDN}, the Seiberg-Witten
differential is defined by finding $\lsw$ so that
$\Omega_{(2)} \equiv \omega \wedge du = d \lsw$, where
$\omega$ is the holomorphic differential on the torus.
The problem is that $\Omega_{(2)}$ is not exact -- it
has non-trivial integrals over the $2$-cycles in the
surface that is defined by $(x,y,u)$.  To define $\lsw$
one must first excise these $2$-cycles, and for $\lsw$ to
be meromorphic, once must excise these curves holomorphically.
One also wants to preserve the proper discrete symmetries,
so one must make excisions in an appropriately Weyl invariant manner.
Thus one is to excise Weyl orbits of rational curves: but
there is the choice of the degree of these curves. One
usually excises lines, but this is for the sake of simplicity
and convenience:  As was evident in \JMDN\ one could
equally well excise quadratics or cubics, or even higher
degree curves.  As described earlier, rational curves
are labelled by weight vectors, and the length of the
weight vector increases with degree.  By Bezout's theorem,
such a curve generically intersects the Seiberg-Witten torus
(defined by $u= constant$) $d$ times.  Thus excising such a
curve introduces in $\lsw$, $d$ simple poles each with the residue
$\vec v \cdot \vec m$, where $\vec v$ is the weight label,
and the components of $\vec m$ are the mass parameters.

Thus excising a Weyl orbit of rational curves of degree $d$
gives rise to $d$ poles for each vector in the orbit, and
hence the indecomposable BPS states come with an additional
degeneracy factor of $d$.  Thus we have another understanding
of the degeneracy of the curves of arithmetic genus $0$ --
the multiplicity comes from the intersections of each curve
with the Seiberg-Witten torus.

One can also begin to see how the degeneracies might
work for curves of non-zero arithmetic genus.
As we stressed in the previous section, the degeneracies
are not simply the Euler number for the moduli space
of rational curves, but probably some combination
of Euler numbers of various pieces of the moduli space.
Here we propose a slightly more precise physical description:
the degeneracy is computed by counting  intersections of
curves of degree $d$ with the planes $u= u_0$, where $u_0$ is a
constant  ({\it i.e.} intersections with the Seiberg-Witten torus).
As described above, each such curve intersects the torus
$d$-times, and so each degeneracy polynomial must
have a factor of $d$.
For higher arithmetic genus the necessary refinement is that
the presence of double points may require that we look
at tori at specific values of $u_0$, and then sum over
choices of $u_0$.  For example, one of the points in the moduli space
of the  curves with arithmetic genus $1$ is  the torus
itself, {\it i.e.} the curve $u = u_0$.
For this curve to be rational, it must have a double point, and
so the discriminant must vanish at $u=u_0$.
For $E_8$  the number of such zeroes is $12$, and for $E_n$ it
is $n + 3$.  For $E_n$, curves of
arithmetic genus $1$ first appear at degree $d = 9 -n$, and
since the number of moduli is $d-1$, the degeneracy for $E_8$ is
simply $12$ (there is a singular fiber at $u=\infty$).
For $E_n$, $n<8$ each singular torus belongs to a family of rational
curves of the same degree.  A generic member of this family intersects
$u=u_0$ at $d$ points.  Thus, counting the self-intersections of the
singular Seiberg-Witten tori gives us  $d(12-d)$ which is the correct
degeneracy for $g_a=1$ rational curves.  We also see more clearly that
the  ``error''
for $d=1$ is associated with the singular fiber at $u=\infty$.

The foregoing discussion of curves  is far from rigorous,
but very suggestive.
It would thus be very interesting to revisit the IIB string
compactifications that lead to these models and see how
the choice of the degree of the excised curves arises in
the construction, and how
the BPS geodesic methods are to be modified so as to
properly incorporate the curves of higher arithmetic
genus.  This might lead to a simple physical understanding
of the degeneracy polynomials and their relationship to
the topology of moduli spaces.

\newsec{Toroidal compactification of the non-critical string}

Thus far we have considered the non-critical string
compactified on a degenerate torus with $R_5/R_6 = \infty$.
We now briefly consider the corresponding story with
$R_5/R_6$ finite.

{}From the form of the trigometric tori in the appendices
and in section 2, it is fairly obvious how to restore the
lost modulus of the torus, and the corresponding double
periodicity of the complexified Wilson line parameters.
For the curves with up to two non-zero masses
one simply replaces the sine functions by the
corresponding Jacobi elliptic functions, $sn(u,k)$,
where $k = \vartheta_2^4(0|\tau)/\vartheta_3^4(0|\tau)$ is the
elliptic modulus, and $\tau$ is the usual Teichmuller parameter of
the torus.   The surface \twomass\ becomes
\eqn\elltwomass{\eqalign{ y^2 ~=~ x^3 &~+~ (1 + k^2) u^2 x^2 ~+~
k^2~u^4~ x \cr &~-~ 2u ~ (u^2 ~+~ sn^2(m_+) x)~
(u^2 ~+~ sn^2(m_-) x) \ .}}
It is shown in Appendix A that this is indeed the proper
form for the curve.  For the curves with more than two non-zero
masses the situation becomes considerably more complicated
(essentially because there are many natural elliptic functions
that reduce to unity in the trigonometric limit).  However,
the curves can be obtained using the approach of \GMS.

In appendix A we also construct the Seiberg-Witten
differential associated with this surface, and even
for two masses the explicit expressions are extremely
complicated.  The important point for the discussion
here is that the logarithms in \indefint\ are replaced
by inverse elliptic functions:
\eqn\invell{ \int_0^{sn(m)} ~{dt \over (1 - t^2)(1 - k^2 t^2)}
{}~=~ m \ .}
This is necessary to make the residues of the Seiberg-Witten
differential linear in the masses, while having the curve itself
doubly periodic in its dependence on the masses.  Moreover, the
non-critical string compactified on
a torus must have Kaluza-Klein excitations of mass
$2\pi i (N_1/R_5 + N_2/R_6)$, for all integers $N_1$ and $N_2$.
One sees that this is
properly encoded in the differential if it has a prefactor
of $1/R_5$ as in section 2, but is now the inverse of
an elliptic function with $Im(\tau) = R_5/R_6$.

We can now push the construction of \LMW\ backwards
so as to reconstruct a non-compact Calabi-Yau manifold
with two moduli that are the complexified versions
of $k_D + k_E$ and $k_E$.  To resolve an ambiguity in how to do
this we need a slightly more explicit form of the Seiberg-Witten
differential.  The construction in Appendix A
generically involves the following indefinite integral:
\eqn\invsnu{ \int_0^1 {dv \over \sqrt{x^3 + (1 + k^2) u^2 v^2 x^2 +
k^2 u^4 v^4 x  + f(x,u;m_i)}} \ ,}
where
\eqn\torpert{f(x,u;m_i) ~\equiv~  y^2 - (x^3 + (1 + k^2)
u^2 x^2 + k^2 u^4 x) }
represents the perturbation of the curve away from the
massless point.  In particular, note that $f$ is independent
of the integration variable, $v$.  Reversing the calculation
of \LMW\ this suggests that we should interpret $v$ as
one of the Calabi-Yau coordinates, and the point
$v=1$ should correspond to a limit of integration that is
set (as in \LMW) by the integration over the third Calabi-Yau
coordinate.  Thus, one can easily arrive at the following
expression for the non-compact Calabi-Yau manifold
(without mass parameters \foot{The construction of
the Calabi-Yau manifold with mass parameters cannot be done
 by such a simple procedure:  if one tries the naive
approach one obtains a Calabi-Yau manifold with too few
independent moduli.}):
\eqn\pertCY{w^2 ~=~  z_1^3 ~+~ z_2^6 ~+~ z_3^6  ~-~
{1 \over z_4^6} ~+~ \psi^2 (1+k^2) (z_1 z_2 z_3 z_4)^2 ~+~
k^2 \psi^4 z_1 (z_2 z_3 z_4)^4 \ .}

Given the identification of the new parameter in terms
of the torus compactification of the non-critical string,
one can use the work of \KMV\ and \LMW\ to relate it
to modulus of the Calabi-Yau in the IIA compactification.
Indeed it is the complexification, $t_E$,
of the K\"ahler modulus $k_E$.

Thus we propose that this non-compact Calabi-Yau manifold
(or the corresponding torus) captures the sector of
the non-critical string that is defined by the closed
sub-monodromy problem described in \LMW\ that associated with
the two parameters $t_S$ and $t_E$.  In practical terms,
this means that we should be able to pass between phase
$I$ ($t_S < t_E$) and phase $II$ ($t_S > t_E$),
perhaps seeing the phase transition or some curve
of marginal stability.  We should also be able to generate
the full instanton counting
of \KMV, with independent $d_E$ and $d_D$ independent.
Work is continuing along these lines, and preliminary
calculations indicate that one should be able to
find expressions for the degeneracies explicitly in
terms of modular functions of $\tau$. The precise computation
is rather complicated as one needs to carefully evaluate
the ``constants of integration,'' in $\phi$, and these
``constants'' can, in principle, be very complicated functions
of $\tau$.

\newsec{Conclusions}

We have shown that the effective action
of a non-critical string does indeed capture
much of the information about the BPS structure
of the theory.  We have shown how the formulation
of the massive effective action is closely parallel
to the corresponding object in field theory, and
yet it contains information that is appropriate
to the string compactification.  In particular we used this
effective action to count BPS states replete with
the full set of $E_n$ character parameters. The fact that
this computation works and provides answers that are
consistent with results from field theory and
from Calabi-Yau manifolds already represents a
remarkable number of consistency checks on the
overall approach.  Combined with the internal
self-consistency of the character expansions, and
the proper behaviour of the flows from $E_8$ to
$E_n$, we feel that we have made a compelling case,
not only for the correctness but also for the utility
of the approach.

We have focussed on a particular sub-sector of the
non-critical string:  One that corresponds to
the string wrapped $d$ times around a circle and
in a phase where all such states are becoming
massless.  The general belief is that such
multiply wrapped strings should be bound states.  By extracting
the degeneracy polynomials in Table 5, we have
obtained the first predictions for the universal ($E_n$
independent) degeneracies of such bound states.
As we remarked in section 5, there is almost
certainly a beautiful mathematical characterization of
these degeneracies.  However, we believe that there should
also be some simple physical description of the degeneracies.

Based upon the
picture of non-critical strings in terms of membranes
stretching between a $9$-brane and a $5$-brane,
a natural suggestion for the spectrum is a represenation of
$E_8$ current algebra at level one (inherited from the
$9$-brane) multiplied by some eta-functions (associated
with the $5$-brane).  This is indeed what one finds \KMV\ for
the lowest level of the non-critical string ($d_E = 1$).
The most natural first guess for the compactified and
multiply wound string is the same current algebra representation,
but with more complicated structure, perhaps through level
matching, coming from the $5$-brane degrees of freedom.
This possibility is ruled out by our data:
The level one representations generally have every
Weyl orbit occurring once.  One might be able to get
around this by multiplying by some eta-functions, or other
modular functions and then doing some exotic level matching, but
our data shows that distinct Weyl orbits of vectors of the
{\it same length} usually come with different degeneracies.
This cannot be realized by a simple level matching of level one
$E_n$ characters with other modular functions.
A less naive suggestion is that the bound states may
involve $E_n$ current algebras at level $d$ for the
string wrapped $d$ times.  Preliminary calculations
for $d=2$ indicate that this does not appear to work either.
Thus, in spite of all this data, a simple
physical characterization
of the spectrum of the $E_n$ string still eludes us.

Another potentially useful physical application of our
degeneracy polynomials is in entropy calculations where
one need to estimate of the growth of the number of
BPS states \CVPC.   For example, the degeneracy polynomials strongly
suggest  that the number of $E_n$ Weyl orbits corresponding to
curves of degree $d$ and arithmetic genus $g_a$ grows as
$d^{g_a+1}/g_a !$.  Summing over $g_a$ for a given $d$
one sees that the number of Weyl orbits must grow as
$e^d \sim e^{L}$, where $L$ is the length of
the vectors in the corresponding Weyl orbit.  Since
the number of weight vectors of $E_n$  of a given length
grows as $L^n$, and so this does not modify the
exponential growth.  A similar result has been
obtained from a numerical fit using the mirror map
on the Calabi-Yau manifold \PMPC.

Finally, we believe that the ideas described in section 6 will
lead to far more complete characters for the states of
the $E_n$ non-critical strings.  In particular, we
expect to obtain explicit modular functions for some of
the non-critical string degenaracies.  This should
not only shed some light on the modular structure
of the spectrum, but should also enable some sharp
estimates of the growth of the number of BPS states.

\goodbreak
\vskip2.cm\centerline{\bf Acknowledgements}
\noindent
We would like to thank S.~Katz, W.~Lerche, P.~Mayr and C.~Vafa for
valuable discussions.  N.W.~is also grateful to the ITP in
Santa Barbara, and the Institute for Advanced Study in
Princeton for hospitality while this work was being done.
This work is supported in part
by funds provided by the DOE under grant number DE-FG03-84ER-40168,
and by the National Science Foundation under grant No. PHY94-07194.

\vfill\eject

\appendix{A}{Derivation of the $E_8$ curve with up to two Wilson lines.}

\def\sp{sn m_+}
\def\s2{sn m_-}
\def\d1{dn m_+}
\def\c1{cn m_+}
\def\coeff#1#2{\relax{\textstyle {#1 \over #2}}\displaystyle}
\def\coeff#1#2{\relax{\textstyle {#1 \over #2}}\displaystyle}

\subsec{Introducing one mass}

To construct the curves with the mass parameters, we follow the
methods developed in \JMDN.   Before starting it is convenient to
make the shift $x\rightarrow x-\coeff1{4}u^2$ in \canoncurve, and
rescale to obtain the curve
\eqn\shifte{  y^2 \ = \ x^3- { 1\over 48 } u^4 x + {1 \over 864} u^6 +u^5 . }
We now add one  mass parameter, which breaks the  $E_8$ symmetry
down to $SO(14)$. The general form of the
curve  consistent with $SO(14)$ symmetry has the form
\eqn\pertcurve{ y^2 \ = \ x^3-({ 1\over 48 } u^4 + bu^3 +
3\lambda u^2 ) \ x \ + \
({1 \over 864} u^6 + \beta u^5 \gamma u^4 +2\lambda u^3) \ .}
Here $ b,  \beta, \ \gamma$ and  $\lambda  $ are constants that we  will
determine bellow.  The discriminant of \pertcurve\  is given by
\eqn\disc{  \Delta \ = \ 4 \ ({1\over 48 } u^4 + bu^3+3\lambda u^2)^3-
 27 \  ({1 \over 864} u^6 + \beta u^5 \gamma u^4 +2\lambda u^3)^2 . }
Since the global symmetry of the pertubed curve is $SO(14)$
the discriminant has to be of order
$u^9$ as $u \to 0$.   This fixes: $\gamma =b \lambda  $ and
$\beta \ =  \ b^2 /12 \lambda$.  The
rest of the constants can be fixed by finding the appropriate lines.
We will assume that the lines have the same form as the ones in the
polynomial curve of \JMDN. Therefore we look for lines of the form
\eqn\lines{ x \ = \ \mu^2 u^2\ + \ \nu u  \ . }
With  $\nu \ = - \lambda $   we   have the spinor line. The adjoint
line is obtained by setting  $ \nu \ = \ 2 \lambda$.
First consider the spinor line.
If we set $\mu \ = \ t^2 - \coeff 16$  it is easy  to verify that
the line \lines\ gives rise to a  perfect square, and one obtains
\eqn\sqy{  y \ = \  i u^3 (t^2-\coeff14) t \ .}
For our later analysis it is convenient to
shift back  $ x\rightarrow  x + \coeff1{12}u^2 -\lambda u$.
The curve with one mass now has the form
\eqn\newfor{ y^2 \ = \ x^3 +(\coeff14 u^2 - \lambda u)x^2 +6 \lambda
(t^2 - \coeff14)u^3 x \ - \ 3 \lambda ( t^2 - \coeff 14)^2 u^5.}
Now it is also easy to verify that adjoint line is
\eqn\adline{  x \ = -(t- {1 \over 4t})^2 u^2 -3 \lambda u}
with
\eqn\ady{ y  \ = \ -~i \Bigl (\coeff18 (t- {1 \over 4t})^2
(t+{ 1\over 4t})^2 u^3-
\coeff 92\lambda^2\ u(t- {1 \over 12 t })\Bigr ) \ .  }
Note here we have written the adjoint line for the curve \newfor.

One can easily recover to the polynomial limit of this curve as
$\lambda \rightarrow  0 $. From \newfor\ we see that $t$ has to
diverge in order to have a finite  $u^5$ term. With $t \approx
\Lambda /m$ and $\lambda \ =\ -\coeff13 \Lambda^2 m^4$  it is
easy to see that  when $m\rightarrow0$ we reduce to the polynomial
limit.

Next we construct the Seiberg-Witten differential $\lambda_{SW}$
for the foregoing curve.  By definition one must have
\eqn\defsw{{ d \lambda_{SW} \over du } \ = \ {dx \over y } +
{d \over dx } ( \ldots) \ .}
One starts by considering the following differential:
\eqn\diffe{  \log \left( { y+ \coeff12 u  x \over y - \half u x}
\right) {dx  \over x }  \ , }
where
\eqn\curve{y^2 \ = \ x^3+ (\coeff14 u^2-3 \lambda )x^2 +
6 \lambda \xi u^3 x -3 \lambda \xi }
and $ \xi \ = \ (t^2 -1/4) $.  The derivative of \diffe\ with
respect to $u$ gives
\eqn\ddife{ { d \over du}  \log \left( { y+ \coeff12 u  x
\over y - \half u x} \right) {dx \over x } \ = {\ x^3 -
\coeff32 \lambda u x^2 - 3 \lambda
\xi u^3 x + \coeff 92 \lambda \xi^2  u ^5 \over y^2 -\coeff14 u^2 x^2 }
\  {dx \over y } \ .  }
This shows that \diffe\ will not do the job and we need additional terms.
First consider  differential
\eqn\addif{{ 1 \over u } {d \over dx }  \log \left( { y + \coeff12
u x \over y- \coeff12 u x } \right) dx \ =  \ { - \coeff12 x^3 +
3 \lambda \xi u^3 x - 3 \lambda \xi^2 u^5 \over y^2 -\coeff14 u^2 x^2}
{ dx \over y }\ . }
It is clear from \addif\ that \ddife\  is not enough
to cancel off the dominator. To this end consider
\eqn\newterm {  \log \left( { y +\coeff12 u\ x \over y - \coeff12 u \ x }
\right) { dx \over l_s}\ ,}
where $ l_s \ = \ x- \xi u^2 $ is the spinor  line.  With this definition
we can rewrite our curve in the form
\eqn\simple{\eqalign{  & y^2 \ = \ l_s \ q + r^2 \ ,  \ \ \ \  r_s \ = \
u^3t\xi \cr & q \ = \ x^2 +(t^2u^2 -3\lambda  u ) x +u^2 \xi (u^2 t^2 +
3 \lambda u )\ .}}
The derivatives of \newterm\  are given by
\eqn\newder{ \Bigl ( {d \over d u } - { d \over d x } { dl \over d u }
\Bigr )
\log \left( { y +\coeff 12 u x \over y - \coeff 12 u x } \right)
{ dx \over l } \ = \ {- \coeff12 \{ l, q \} u \ x + x q -uq
\coeff{dl}{dq} - 2 r^2\over y^2 -\coeff14 u^2 x^2 } { dx \over y }\ ,}
where $\{ l,q\} \ = \
\coeff{dl}{dx}\coeff{dq}{du}-\coeff{dl}{du}\coeff{dq}{dx}$. Combining
equations \ddife , \addif\ and \newder\ we can construct a Seiberg-Witten
differential
\eqn\swspi{\lambda^{(s)}_{SW} \ = \   \log \left(  { y + \coeff12  u x
\over y - \coeff12 u x } \right) {dx \over x } - \coeff23
\log \left( { y + \coeff12  u x  \over y - \coeff12 u x } \right)
{dx \over x - \xi u^2 } }
that satisfies
\eqn\eqsw{\eqalign{&{d \over du }\lambda^{(s)}_{SW}\ = \ \coeff16
{dx \over y }- \coeff13 { d \over dx }
\Bigl(   \log \left(  { y + \coeff12  u x  \over y - \coeff12 u x }
\right) {dx \over u } \Bigr )-  { d \over d x } 2\xi u \   \log
\left( { y + \coeff12  u x  \over y - \coeff12 u x
} \right) {dx \over x - \xi u^2 }   \ .}}
In the this construction we have only used  the spinor line, but
the full Seiberg-Witten differential should also contain a contribution
from the adjoint line
\eqn\adline{ l_a= x -(t- {1\over  4t})^2u^2 - 3 \lambda u    \ \ \
r_a \ = \ \coeff18 (t- {1 \over 4t})^2(t+{ 1\over 4t}) u^3- \coeff92
\lambda^2\ u(t- {1 \over 12 t })\ .}
In order to include
the adjoint line $l_a$ we have to proceed in a slightly different
fashion. Note the argument of  $  \log \left({ y + \coeff12
(ul_s+\alpha_sr_s)  \over y - \coeff12(ul_s+\alpha_s r_s)} \right)$
is different from   $\log \left({ y + \coeff12(ul_a+\alpha_s r_a)
\over y - \coeff12  (ul_a+\alpha_s r_a)} \right)$.
This means if one takes derivatives of these functions with respect
to $u$ the dominators will generically be different, and so
these terms cannot combine to form a Seiberg-Witten differential.
However if we choose $\alpha_s$ and $\alpha_a$ so that
\eqn\cond{ -x_s u +
\alpha _s r_s \ = \ - x_a u + \alpha _a r_a }
then the dominators will be the same and we can construct a
Seiberg-Witten differential.
We can solve equation \cond\ for $\alpha_s$ and $\alpha_s$ by comparing
the terms  of order  $u^2$ and $u^3$. After simple algebra we have
\eqn\sol { \alpha_s\ = \ - \alpha_a \ = \ { 2 \over 3
(t-\coeff1{12 t} )}\ .}
It is easy to see that  the Seiberg-Witten differential
with  the adjoint line will be of the form
\eqn\swad{ \lambda_{SW}^{(a)} \ = \  \log \left({ y + \coeff12(ul_a +
\alpha_a r_a) \over y - \coeff12(ul_a+\alpha_a r_a)} \right) \Bigl(
{dx \over l_a+ \alpha \coeff{r_a }u}+ b_1 \ {dx \over l_a} + b_2
{dx \over l_s} \Bigr ) \ , }
where  the constants $b_1$ and $b_2$ are determined by demanding that $
\lambda_{SW}^{(a)}$  satisfies the following condition
\eqn\condsw{ {d \over du }\lambda_{SW}^{(a)} + {d \over dx} (
\cdots ) \ = \ k {dx \over y }\ .}

A simple substitution allows as to solve  for the unknown coefficients
to obtain $ b_1 \ =\ -\coeff12 , \ b_2 \ = -1 $ and $ k\ = \ 0 $.
Surprisingly, we find $k=0$, and so we have a {\it null} contribution
to the Seiberg-Witten differential.
It is clear that the full Seiberg-Witten will be a linear combination
of $\lambda^{(s)}$ and $\lambda^{(a)}$
\eqn\fullsw{ \lambda_{SW} \ = \ C_1 \lambda_{SW}^{(s)} \ + \  C_2
\lambda_{SW}^{(a)}\ .}
We fix the coefficients $C_1$ and $C_2$ by  comparing this  with the
superconformal or polynomial limit.  We described above how to make
this limit for the curve, and applying the same procedure to the
Seiberg-Witten differential, we find that the leading term is
given by
\eqn\limsw{({C_1 \over 3} - {C_2 \over 2 }) u {dx \over y } -
( \coeff23 C_1+\coeff23){ r_s \over t } { dx \over y l_s}
+\coeff13 C_2 {r_a \over t } {dx \over y \ l_a}\ .}
The Seiberg-Witten differential in the polynomial limit
with one mass is given by
 \eqn\sewsup{ {1 \over 2 {\sqrt 2} \pi} ( 60 u \ { dx \over y
}-   64 i {\coeff{m}2 r_s \over x-x_s } {dx \over y }-  14 i
{m r_a \over x- x_a} {dx \over y } +42 u {dx \over y})\ .}
They are consistent if we set $C_1 \ = \ { 180 \over 2
{\sqrt 2} \pi }$ and $C_2 \ = \ { -84 \over 2 {\sqrt 2} \pi } $
and then we have
\eqn\finalsw{ {d \over du } ( C_1 \lambda_{SW}^{(s)} \ + \  C_2
\lambda_{SW}^{(a)} ) \ = \ { 30 \over 2 \sqrt2 \pi } {dx \over y }}
Where the coefficient, $30$, is the $E_8$ dual Coexeter number,
as it should be.

To obtain a physical interpretation of the parameter $t$ one
calculates the residue both for the spinor and adjoint line.  The
residue of the spinor line is $ \log{t+\coeff12 \over t-\coeff12}$
and the residue of the adjoint is $ 2 \log{t+\coeff12 \over t-
\coeff12}$. Since the adjoint has a residue that is twice the residue
of the spinor we can identify the residue with the mass term
\eqn\mass{  \log{t+\coeff12 \over t-\coeff12} \ = \ m \ .}
Before generalizing this result to  two masses it is useful to
summarize the final form of our curve with one mass. We have found
that curve with one mass is given by
\eqn\onemass{ y^2 \ = \ x^3 + {u^2  \over 4 } x^2 -2 \Lambda ^6 u
(4 \sin^2 { m \over 4 \Lambda } x+u^2 )^2 \ .}
Here we have set $\lambda \ = \ \coeff{32}3 \Lambda^6 \sin^4
\coeff{m}{4\Lambda}$.

\subsec{The curve and differential with two masses}

We start the case of two masses by first studying the polynomial limit
\eqn\curvecon{ y^2 \ = \ x^3 - (T_2 u^3 + {\tilde   T_4^2 u^2
\over 12 } ) x- (2u^5 +u^4 { T_2\tilde T_4 \over 6 } +
{ \tilde T_4^3 \over 108  } )\ , }
where $ \tilde T_4 \ = \ \coeff14 T_2^2 - T_4 $ and $T_2 \ = \ m_1^2 +
m_2^2 $,  $ T_4 \ = \ m_1^2 m_2^2 $.
After shifting $x \rightarrow x - \coeff16\tilde T_4 u$ the curve becomes
\eqn\shift{\ y^2 \ = \ x^3 - 2 \ u x^2 ( {m_1+m_2  \over 2} )^2
( {m_1-m_2  \over 2} )^2- 2 u^3 x (({m_1+m_2 \over 2 })^2+
({m_1-m_2 \over 2 })^2) - 2 \ u^5}
In the shifted form the spinor line has a very simple form
\eqn\lines{x \ = \  { 4 \over (m_1 \pm m_2 )^2 } \ u^2}
with
\eqn\valy{y \ = \ {8 i \over (m_1 \pm m_2)^2 }u^3 }
The adjoint lines, $l_\pm$,  have the form
\eqn\anline{ x \ = \ - {1 \over ( m_+\pm m_- )^2}u^2 + 2 m_+^2m_-^2 \ ,}
with
\eqn\vally{ y \ = \ i( {u^3 \over (m_+ \pm m_-)^3 }\pm {2 \ m_+ m_-
\over m_+ \pm m_- } ( m_+^2 \pm m_+m_- + m_-^2 )~u^2) \ . }
Here we have also introduced the notation $ m_{\pm} \ = \ \coeff
{m_1\pm m_2}2$ .

To construct the trigonometric curve we make the Ansatz
\eqn\curt{  y^2 \ = \ x^3 +\coeff14 u^2 x^2 + \coeff12
T^{\prime}_4 u x^2- 2 \ u ( u^2 +   \coeff14  T^{\prime}_2 x) \ . }
Recall that with one mass the curve was obtained by replacing $m^2$
with $16 \sin^2 \coeff {m}4 $.  This suggest then that  the curve
with two masses  is obtained in a very similar fashion, namely
replacing $ m_{\pm}$ with $ 4 \sin^2\coeff{ m_{\pm}}4$. This
then leads us to the curve
\eqn\curtwo{ y^2 \ = \ x^3 + \coeff14  u^2 \ x^2 - 2 u (4 x  \ \sin^2 \
\coeff{m_+}2 + u^2 ) (4 x  \ \sin^2 \ \coeff{m_-}2 + u^2 ) \ .}
Again we look for lines, starting with the spinor line.
The generalization of equation \lines\ is straightforward and we have
\eqn\genline{ x \ = \ - { u^ 2 \over 4 \sin ^2  \coeff{m_{\pm}}2} \ ,}
with
\eqn\newyva{ y \ = \ i {  \cos~m_{\pm} /2 \over 8 \sin^3 m_{\pm}}\ .}
Next we consider the adjoint line that has the form
\eqn\secline{ x \ = \ - {u^2 \over 4  \sin^2 \coeff { m_++m_-}2 }  +
32 \sin^2 {m_+ \over 2}\sin^2 {m_- \over 2} \ u \ ,}
with
\eqn\yeq{\eqalign{ y \ &= \  {  i \ u^3  \cos { m_+ + m_- \over 2 }
\over 8 \ \sin ^3 { m_+ + m_- \over 2}} \pm { 16 \ i \ u^2
\sin { m_+  \over 2 }\sin { m_- \over 2 }\over \sin { m_+ +m_-
\over 2 }}\cdot  \cr & ( \sin^2 { m_+ \over 2 }+ \sin^2
{m_+ \over 2 }  -\sin^2{ m_+ \over 2 }\sin^2{ m_- \over 2 } \pm
\sin{ m_+ \over 2} \sin{ m_- \over 2}  \cos{ m_+ \over 2}
\cos{ m_- \over 2}) \ . }}
As before, in order to construct the Seiberg-Witten differential
we start with the spinor residue and rewrite our curve using
the spinor line $l_s$ in the form $y^2 \ = \ l_s~q +r^2_s$ where
\eqn\rql{\eqalign{  l_s \ & = \ x+ { u^2 \over 4 \sin \coeff{ m_{\pm}}2}
\ , \ \ \ r_s ^2 \ = \ - {  \cos^2\coeff{ m_{\pm}}2 \over 64 \sin^6
\coeff{ m_{\pm}}2 }u^6 \cr q \ & = \ x^2 - { \cos \coeff{ m_{\pm}}2
x \over 4 \sin^2 \coeff{ m_{\pm}}2}u^2 + { \cos^2\coeff{ m_{\pm}}2
\over 16 \sin^4\coeff{ m_{\pm}}2}u^4 - 8usin^2 \coeff{ m_{\pm}}2(u^2 +
4 \sin^2 \coeff{ m_{\pm}}2 x ) \ .}}
Consider the differential
\eqn\swdif{ \log \left( {y + \coeff12(ul+\alpha r) \over y + \coeff12
(ul+\alpha r) } \right) {dx \over l }\ .}
If we set $ \alpha =\alpha_{\pm}= {2i \sin \coeff{ m_{\pm}}2 \over
\cos \coeff{ m_{\pm}}2}$ it is easy to see that  the residue $ im_{\pm}$.
Proceeding in the same fashion as in the one mass case one can construct a
Seiberg-Witten differential using the spinor line:
\eqn\spisw{{d \over du }\Biggl (  \log \left(  { y + \coeff12 ux
\over y -  \coeff12 ux } \right) \Bigl ( {dx \over x } - \coeff13
{dx \over l_+} -  \coeff13 { dx \over l_-}\Bigr )\Biggr ) \ = \
\coeff16 {dx \over y } +  \ { d\over dx}( \ldots ) \ .}
With only one mass we saw that the full Seiberg-Witten differential
requires the existence of a null differential.  It is not surprising
that this will also be also for two masses.  Again before solving for
these null differentials it is instructive to study the polynomial
limit.  For two masses we have are six different poles:
$ m_+,\ m_- , \ m_+ +m_-, \ m_+-m_-, \ m_{a_+}=2m_+$  and
$ m_{a_-}=2m_-$. As we have seen, the spinor lines correspond to
residues $m_\pm$ and the adjoint lines corresponds to the residue
$ m_+\pm m_-$. The residues $2 m_\pm$ can be
identified with a second set of adjoint lines $l_{a_\pm}$
\eqn\neadjoint{ x \ =\  { -u^2 \over 4 m_\pm^2}+u
(3m^2_\pm-m^2_\mp)m_\pm^2-m_\pm^6(m^2_+-m^2_-)^2 \ ,}
with
\eqn\valyap{ y^2 \ = \ - ( {u^3 \over 8 m_\pm^3}+ 6 u^2 (5
m_\pm^6-m_\pm^4m_\mp^2)-4um_\pm^8(9m_\pm^2-5m_\mp^2)
(m_\pm^2-m_\mp^2)+8m_\pm^{12} (m_\pm^2-m_\mp^2)^3 )^2 \ .}
It turns out that one can make a consistent Seiberg-Witten
differential  out of any two lines $l_a$ and $l_b$. The following Ansatz
leads to a Seiberg-Witten  differential:
\eqn\ans{\eqalign{ & {d \over du } i m_a r_a {dx \over l_a y} -
{dl_a \over du } {d\over dx} i m_a r_a { dx \over l_a y }+ A_b\Bigl
( {d \over du } - {dl_b \over du
} {d \over dx} \Bigr ) i m_b r_b {dx \over l_b y }+ {d \over du }
\alpha (U+T) {dx\over y }\cr &
+ \beta {d \over dx} (x +S ) {dx \over y } \ = \ k { dx \over y } \ .}}
For example if we  choose  the lines $ l_+$ and $l_-$ we find  that the
unknown coefficients are given by $ A=-1 $,
$k=-\coeff12$, $T=0$, $\beta=1$ and $S = 2u^2 (m_+^2+m^2_-)/m^2_+m^2_-$

We can now use this this result find the final set of adjoint lines. The
missing lines are the generalization of the adjoint lines $l_{a_\pm}$
of the polynomial curve. From our experience with the one mass case one
should be able to construct
a null Seiberg-Witten differential using the lines $l_+$ and $l_{a_\pm}$.
If this is the case then we have to satisfy the following condition
\eqn\condo{  \log {y+\coeff12 (ul_+ + \alpha_+r_+ + l_+ C) \over
y-\coeff12 (ul_+ + \alpha_+r_+ + l_+ C) } \ = \
\log  {y+\coeff12 (ul_{a_+} + \alpha_{a_+}r_{a_+} + l_{a_+} C) \over
y-\coeff12 (ul_{a_+} + \alpha_{a_+}r_{a_+} + l_{a_+} C) } \ . }
For $l_+ = x + u^2/4\sin^2\coeff{m_+}2 , \ r_+ = iu^3 /8 \sin^3
\coeff{m_+}2 $ and for  $ l_{a_+} = x+\coeff14 { u^2 \over
\sin \coeff{m_+}2} -b_1u-\lambda^2$ and $
r_{a_+} \ = \ {i \cos m_+\over 8 \sin^3 m_+}u^3 + a_2 u^2 a_1u \lambda^3$.
Substituting these into \condo, and demanding the $l_+$ is line on our
curve, gives, after some algebra:
\eqn\sol{\eqalign{\lambda \ & = \ -32 i{ \sin ^3 \coeff{m_+}2 \over \cos
\coeff{m_+}2  } (\sin^2 \coeff{m_+}2 -\sin^2\coeff{m_-}2) \cr
 b_1 \ & = \ { 16 \sin^2 \coeff{m_+}2 \over \cos^2\coeff{m_+}2 }\Bigl(
 \coeff12 \sin^2 m_+ \sin^2\coeff{m_+}2 -\sin^2\coeff{m_-}2\Bigr )\cr
a_1 \ & = \ {-256i \sin^5\coeff{m_+}2 \over \cos^3 \coeff{m_+}2} (\sin^2
\coeff{m_+}2-\sin^2\coeff{m_-}2)\Bigl( 3 ( \sin^2 \coeff{m_+}2-\sin^2
\coeff{m_-}2) +2 \cos^2 \coeff{m_+}2(3 \sin^2\coeff{m_+}2-\sin^2
\coeff{m_-}2 )\Bigr )\cr a_2 \ & = \ 8i (\cos~ m_+ +2 ){ \sin^3
\coeff{m_+}2 \over \cos \coeff{m_+}2}+ 6i
{\sin \coeff{m_+}2 \over \cos^3 \coeff{m_+}2}( \sin^2
\coeff{m_+}2-\sin^2\coeff{m_-}2)\cr
C \ & = \ -128 {\sin^4 \coeff{m_+}2 \over \cos~m_++ 2}(\sin^2
\coeff{m_+}2- \sin^2\coeff{m_-}2) \ .}}
Having found the second adjoint line we are now ready write down the null
Seiberg-Witten differentials. After some algebra we find the following
null-differentials:
\eqn\nullsws{\eqalign{ \lambda^{(n)}_1 \ & = \  \log \Bigl (
{ y + \coeff12( ul_1 + \alpha_1 r_1) \over y - \coeff12( ul_1 +
\alpha_1 r_1)}\Bigr ) \Bigl ( {dx \over l_1 +\alpha_1 r_1 /u}- \coeff12
{dx\over l_1}- \coeff12 {dx \over l_+}- \coeff12 {dx \over l_-}\Bigr)\cr
\lambda^{(n)}_2 \ & = \  \log \Bigl ( { y + \coeff12( ul_2 +
\alpha_2 r_2) \over y - \coeff12( ul_2 +
\alpha_2 r_2)}\Bigr ) \Bigl ( {dx \over l_2 +\alpha_2 r_2 /u}- \coeff12
{dx\over l_2}- \coeff12 {dx \over l_+}- \coeff12 {dx \over l_-}\Bigr)
\cr \lambda^{(n)}_+ \ & = \  \log \Bigl ( { y + \coeff12( ul_{a_+} +
\alpha_{a_+} r_{a_+}+l_{a_+}C_+) \over y - \coeff12( ul_{a_+} +
\alpha_{a_+} r_{a_+}+l_{a_+}C_+)}\Bigr ) \Bigl ( {dx \over l_{a_+}
+\alpha_{a_+} r_{a_+} /u}- \coeff12{dx\over l_{a_+}}- \coeff12
{dx \over l_{a_+}}-  \coeff12 {dx \over l_+}\Bigr)\cr
\lambda^{(n)}_- \ & = \  \log \Bigl ( { y + \coeff12( ul_{a_-} +
\alpha_{a_-} r_{a_-}+l_{a_-}C_-) \over y - \coeff12( ul_{a_-} +
\alpha_{a_-} r_{a_-}+l_{a_-}C_-)}\Bigr ) \Bigl ( {dx \over l_{a_-}
+\alpha_{a_-} r_{a_-} /u}- \coeff12{dx\over l_{a_-}}- \coeff12
{dx \over l_{a_-}}- \coeff12 {dx \over l_-}\Bigr) \ .\cr}}
The full Seiberg-Witten differential is a linear combination of the spinor
Seiberg-Witten differential and the null Seiberg-Witten differentials
constructed above:
\eqn\linsw{ C_s \lambda_s + C_1 \lambda_1^{(n)}+ C_2 \lambda_2^{(n)}+
C_{a_+} \lambda_{a_+}^{(n)}+ C_{a_-} \lambda_{a_-}^{(n)} \ ,}
where $ C_s=-{180 \over 2 \sqrt2 \pi } $.
This is the same constant that appeared in the one mass case. There are
12 $l_1$  and 12 $l_2$ lines, each have a residue $ m_++m_-$ and
$ m_+-m_-$ and
\eqn\css{ C_1 \ = \ {-24 (m_++m_-)i \over 2 \pi\sqrt2  \log \bigl({ 1+
\alpha_1/2 \over 1- \alpha_1/2}\bigr ) } \ , \ \ \ \  C_2 \ = \
{-24 (m_+-m_-)i \over 2 \pi\sqrt2  \log \bigl({ 1+ \alpha_2/2
\over 1- \alpha_2/2}\bigr ) }\ .}
There is only one $l_{a_+}$ line and one $l_{a_-}$ line with residues
$2 m_+$ and $2m_-$
\eqn\cps{ C_{a_+} \ = \  \ {-2 m_+i \over 2 \pi\sqrt2  \log \bigl({ 1+
\alpha_{a_+}/2 \over 1- \alpha_{a_+}/2}\bigr ) }\ , \ \ \   C_{a_-} \ = \
\ {-2 m_+i \over 2 \pi\sqrt2  \log \bigl({ 1+ \alpha_{a_-}/2 \over 1-
\alpha_{a_-}/2}\bigr ) }\ .}
The total residue of the lines $l_{\pm}$ is $32 m_{\pm}$, as expected
since there are 32 such lines on top of each other.

\subsec{The curve with elliptic parameters}

Next we want to generalize the curve to the elliptic case.
We will start with two masses and take as our Ansatz
\eqn\snucur{ y^2 \ = \ x^3 +  \gamma x^2 u^2 - 2 u \mu (u^2 +sn^2m_+
x)(u^2+sn^2 m_- x) + \beta x u^2\ ,}
where $\gamma,\mu $ and $\beta$ are constants to be determined.
Here we have chosen to write the curve in terms of the
Jacobi elliptic functions
\eqn\ellipticf{ sn(u) \  = \  {\vartheta_3(0) \over \vartheta_2(0) }
{\vartheta_1(u) \over \vartheta_4(u) } \ , \ \ \ \
cn(u) \   = \  { \vartheta_4(0) \over \vartheta_2(0) }
{\vartheta_2(u) \over \vartheta_4(u) }  \ , \  \ \ \
 dn(u) \  = \  {\vartheta_4(0) \over \vartheta_3(0) }
 {\vartheta_3(u) \over \vartheta_4(u)} \ ,}
with
\eqn\defk{ k \  = \ {\vartheta_2^2(0) \over \vartheta_3^2
(0) }\ .}
The reason for using the Jacobi functions, rather than
Weierstrass or theta functions, is that some of our
results above can be easily generalized by just
replacing the trigonometric functions with Jacobi elliptic functions.
We would like to draw the readers attention to  the new term $xu^4$
in \snucur\ since this will  give rise to elliptic functions in the
Seiberg-Witten differential. It is easy to verify that he trigonometric
curve is obtained in the limit of $k \rightarrow  0 $.
To fix the unknown coefficients  we look for lines. As before
we start with the spinor line, which is given by
\eqn\snuspi{ x \ = \ { -u^2 \over  sn^2 m_{\pm}} \ .}
If we set $ \gamma \ = \ 1 +k^2$ and $\beta  \ = \ k^2$ it is easy to
see that equation \snuspi\ gives rise to a perfect square
\eqn\valysnu{  y \ = \ { -i u^3  cn( m_{\pm}) dn (m_{\pm})\over
sn^3(m_{\pm})}\ .} Next we consider the adjoint line of  the form
\eqn\adli{ \ x \ = \  { - 1 \over sn^2 {m_+\pm m_-}}u^2 + b u  \ \ \ , }
where $b$ is  a constant that we determine by  substituting the
line into our curve and demanding that $y$ is a perfect square.
Upon substitution, the lowest power of $u$ in $y^2$ is $u^3$, which
cannot be part of a perfect square, and so setting this to zero
gives:
\eqn\valub{ b \ = \  - 2 \mu sn^2(m_+)~sn^2(m_-)\ .}
With the adjoint line we can  fix the remaining constant in
the curve.  After a rather lengthy but straightforward calculation
we find that if $\mu \ = 1, $ $y$ is a perfect square given by
\eqn\spiy{ y \ = \ { i cn(m_+ \pm m_-) dn(m_+ \pm m_-) \over
cn^3(m_+ \pm m_-) }u^3\ .}

So far we have found the spinor line $l_s$ and the adjoint line $l_\pm$.
 From our previous analysis we know that we have a second adjoint line
$l_{a_\pm}$.  To find this  line we use the same trick as we used
in the trigonometric case, namely set
\eqn\snutrick{ u l_+ + C l_+ \alpha_+ + r_+ \ = \   u l_{ a_+} +
C l_{a_ +}+ \alpha_{a_+} r_{a_+} \ ,}
where
\eqn\deflp{ l_+ \ = \ x \ + { u \over sn m_+} \ , \ \ \ \ r_+ \ = \
{i u^3 cn m_+ dn m_+ \over sn^3 m_+} \ .}
We look for a solution that has the form
\eqn\adjlinep{ l_{a_\pm} \ = x - ( b_2u^2 +b_1u +\lambda^2) \ , \ \ \ \
r \ = \ a_3 u^3 +a_2u^2+a_1 u +\lambda^3 \ .}
Matching the coefficient of different powers of  $u$ in
equation \snutrick\ we find after some algebra
\eqn\avalues{\eqalign{ & a_1 \ = \ 3 sn^2(m_+)( sn^2(m_+) -
{ sn^2(m_-)\over 3} + {sn^2(m_+)-sn^2(m_-) \over
cn(2m_+) dn(m_+)} - i { sn^3(m_+)(sn^2(m_+) -sn^2(m_-))
\over cn^2(m_+) dn^2(m_+) } \cr & a_2 \ = \ i~{cn(2m_+)~
dn(2m_+) \over sn^3(2m_+)}\cr & a_3 \ = \  {i \Delta ( cn^2(m_+)-
sn^2(m_+) dn^2(m_+)) (dn^2(m_+)-k^2 sn^2(m_+) cn^2(m_+))
\over sn^3(m_+)cn^2(m_+)dn^2(m_+)}\cr & b_1 \ = \  2 sn^2(m_+)
\Bigl ( sn^2(m_+) + { sn^2(m_+) -sn^2(m_-) \over cn(2m_+)~dn(2m_+) }
\Bigr )\cr & b_2 \ = \ -{ \Delta^2 \over 4 sn^2(m_+) cn^2(m_+) dn^2(m_+)}
\cr & \lambda \ = \ - i { sn^3(m_+)(sn^2(m_+)-sn^2(m_-) \over cn(m_+)~
dn(m_+) }\cr & \Delta \ = \ 1- k^2 sn^4(m_+)\ .}}

\subsec{Seiberg-Witten differential for the elliptic case}

Having found all the lines  we are ready to construct the Seiberg-Witten
differential.  However, it is first instructive to reconsider the
trigonometric problem. Recall that the Seiberg-Witten differential
has a piece that is  of the form
\eqn\swpart{ \coeff12  \log \left(  {y +ul+ \alpha r  \over y -(ul+
\alpha r) } \right) {u \ dx  \over ul+\alpha r}}
The generalization  to the elliptic case relies on the integral
representation of the log-function
\eqn\intrep{\int^1_0 { u \ dt  \over  (y^2 - (ul +\alpha r )^2 +
(ul +\alpha r )^2 t^2)^\coeff12}}
Similarly the differential $  \log { y + ul + \alpha r \over y- ul -
\alpha r}$ has the following integral representation
\eqn\newintrep{\int^1_0 {ul +\alpha r \over l } { dt \over (y^2 -
(ul +\alpha r )^2 +  (ul +\alpha r )^2 t^2)^\coeff12}\ .}
The generalization of  \newintrep\  to elliptic case is  now
straightforward
\eqn\snuint{\int^1_0 {ul +\alpha r \over l } { dt \over (y^2 - \gamma
(ul +\alpha r )^2 + \gamma (ul +\alpha r )^2 t^2+k^2u^3(u l +\alpha r )
+k^2 u^3 (u l +\alpha r ) t^4)^\coeff12}\ .}
First we can check that that residue at line $ l = x + u^2 /sn^2 (m)$
with  $ r(u)= i u^3 sn(m) dn(m) /sn^3(m) $ and
$ \alpha \ = \ i sn(m)/sn(m) dn(m)$  is indeed  $m$.  From \snuint\ it
follows that the residue at the pole is proportional  to
\eqn\snuresi{ \alpha r \int^{sn(m)}_0 {dt \over u^3 ( 1- \gamma t^2 +k^2
t^4)^{\coeff12}} \ = -i m \ . }
The last equality follows from the definition of the inverse of the
elliptic function.  To construct the Seiberg-Witten differential we
will use the same combination of derivatives that lead to the
Seiberg-Witten differential in the trigonometric case.
Here we want to be left a rational function after performing the
integral.  This will restrict the allowed form of the integrand.
It is easy to see that the integral
\eqn\nonrat{ \int_0^1 {dt  \over (a_0 + a_2 t^2 + a_4 t^4 )^{\coeff12}}}
is not rational, but the following integral is:
\eqn\ratint{\int^1_0 {dt (a_0-a_4t^4)  \over (a_0 + a_2 t^2 + a_4 t^4
)^{\coeff12}} \ .}
We can use this since
for us $ a_0= y^2 - \gamma(ul+\alpha r)- k^2 u^3 (ul+\alpha r)$, $a_2 \ =
\  \gamma (ul+ \alpha r)^2$, $ a_4 \ = \ k^2 u^3 ( ul + \alpha r )$, so
that $ a_0+a_2+a_4\ = y^2 $.
This means that if we can arrange that the numerators is proportional to
$ y^2 - \gamma(ul+\alpha r)- k^2 u^3 (ul+\alpha r) -k^2 u^3 (ul +
\alpha r t^4)$, then after integrating
\ratint\ we get the holomorphic differential.

As before act with the ${d \over du  } - {d l\over du } {d \over dx }$ on the
integral \snuint\ to give
\eqn\inli{\eqalign{ L_1 \ =   \ & ({d \over du  } - {d l\over du }
{d \over dx })\int^1_0 {ul +\alpha r \over l } { dt \over (y^2 -
(ul +\alpha r )^2 + \gamma (ul +\alpha r )^2 t^2 +k^2 u^3 (u l+ \alpha r )
t^4)^\coeff12} \ =\cr & \int^1_0 dt { x^3 +4u^5 +2u^3 + \s2^2 +u^3 x
\sp^2 -\sp^2\s2^2 u x^2)-t^4 k^2 u^4 x \over (y^2 - (ul +\alpha r )^2 +
\gamma (ul +\alpha r )^2 t^2 +k^2 u^3 (u l+ \alpha r ) t^4)^\coeff32}\ .}}
Above  we have used the fact that for a spinor line $u \coeff{dr}{du}= 3r$.
Next consider
\eqn \intlu{\eqalign{ L_u \ = & \ {d \over du }( u \int^1_0 dt
{ 1 \over (y^2 - (ul +\alpha r )^2 + \gamma (ul +\alpha r )^2 t^2 +
k^2 u^3 (u l+ \alpha r ) t^4)^\coeff12})\cr & \int^1_0{ x^3 +
3u^5 +u^3x(\sp^2+\s2) -ux^2\sp^2\s2-k^2t^4u^4x \over (y^2 -
(ul +\alpha r )^2 + \gamma (ul +\alpha r )^2 t^2 +k^2 u^3 (u l+
\alpha r ) t^4)^\coeff32}\ ,}}
and similarly
\eqn \intlx{\eqalign {L_x \ = & \ {d \over dx }( x \int^1_0 dt
{ 1 \over (y^2 - (ul +\alpha r )^2 + \gamma (ul +\alpha r )^2 t^2 +
k^2 u^3 (u l+ \alpha r ) t^4)^\coeff12})\cr  & \int^1_0{ -\coeff12x^3 -
2 u^5 -u^3x(\sp^2+\s2) + \coeff12ux^2\sp^2\s2-k^2t^4u^4x \over (y^2 -
(ul +\alpha r )^2 + \gamma (ul +\alpha r )^2  t^2 +k^2 u^3 (u l+
\alpha r ) t^4)^\coeff32}\ .}}
Consider the
following combination
\eqn\comb{\eqalign{ L_1+L_2-3L_u-L_x \ &  = \ - \coeff12 \int^1_0 dt ({
y^2- \gamma(ul+\alpha r )^2 t^2+ k^2 u^3(ul+ \alpha r) t^4-t^4 k^2 u^4 x
\over(y^2 - (ul +\alpha r )^2 + \gamma (ul +\alpha r )^2  t^2 +k^2 u^3
(u l+ \alpha r ) t^4)^\coeff32 } \cr & = - \coeff12 \Biggl |^1_0
{ t \over (y^2 - (ul +\alpha r )^2 + \gamma (ul +\alpha r )^2  t^2 +
k^2 u^3 (u l+ \alpha r ) t^4)^\coeff1 2 } \cr  \ & = \ -\coeff12 y \ .}}
Hence this combination leads a Seiberg-Witten differential.
The construction of the null differential is very similar to
the trigonometric case. Recall that we had
\eqn\nulleq{\eqalign{ & \Bigl ({d \over du }- {d \over dx }
{d l_s \over du }\Bigr
) \log \left( {y +(ul+ \alpha r )\over y -(ul+ \alpha r
) }\right) { dx \over l_s }+ \coeff12 \Bigl ({d \over du }-
{d \over dx } {d l_a \over du
}\Bigr ) \log \left( {y +(ul+ \alpha r )\over y -(ul+ \alpha r
) } \right) { dx \over l_a }\cr & - \Bigl ({d \over du }u- {d \over dx }
{d u l_a +\alpha r \over du }\Bigr ) \log \left( {y +(ul+ \alpha r )
\over y -(ul+ \alpha r ) } \right) dx  \ = \ 0 \ .}}
This can now easily  generalized to the elliptic case. All we have
to do is to replace the log terms with the appropriate integrals.
The full Seiberg-Witten differential is  again given by a linear
combination of null-differentials and the one constructed out of the
spinor line.

\appendix{B}{General $E_n$ curves}

The polynomial $E_8$ curve is given as
\eqn\conformal{\eqalign{
y^2=&x^3-x^2\bigl(u\widetilde T_4/2+T_{10} - t_8 T_2 \bigr)\cr
& -
x\Bigl(T_2 u^3+ (14 t_8 + T_8) u^2+
  u (8 T_{14} - T_{12} T_2 + 8 t_8 T_6 -T_{10} \widetilde T_4 + 4 t_8 T_2
\widetilde T_4)
 +2 t_8 T_{12}\cr
& + 4 T_{14} T_6
 - T_{12} T_8  +
  2 t_8^2 \widetilde T_4 + 2 T_{14} T_2 \widetilde T_4 - t_8 T_8
\widetilde T_4 + (T_{12} \widetilde T_4^2)/4 + (t_8 \widetilde T_4^3)/4
\Bigr)\cr
&- 2 u^5- T_6 u^4+
  u^3 (4 T_{12} - 2 t_8 T_2^2 - 5 t_8 \widetilde T_4 + (T_8 \widetilde T_4)/2)
\cr
& +
 u^2 \bigl(16 t_8 T_{10} - 8 t_8^2 T_2 - T_{14} T_2^2 + 2 T_{12} T_6 -
     4 t_8 T_2 T_8 - 4 T_{14} \widetilde T_4 + (T_{12} T_2 \widetilde T_4)/2\cr
&
 - 2 t_8 T_6 \widetilde T_4 -
     (T_{10} \widetilde T_4^2)/4 - (t_8 T_2 \widetilde T_4^2)/4\bigr)\cr
& +
  u\bigl (-8 t_8^3 - 2 T_{12}^2 + 8 T_{10} T_{14} - 4 t_8 T_{14} T_2 +
     8 t_8 T_{10} T_6 - 8 t_8^2 T_2 T_6 + 8 t_8^2 T_8 - 2 T_{14} T_2 T_8\cr
& -
     2 t_8 T_8^2 - 3 t_8 T_{12} \widetilde T_4 + 4 t_8 T_{10} T_2
\widetilde T_4 - 4 t_8^2 T_2^2 \widetilde T_4 -
     2 T_{14} T_6 \widetilde T_4 - (T_{12} T_8 \widetilde T_4)/2 - 3 t_8^2
\widetilde T_4^2\cr
& - (T_{14} T_2 \widetilde T_4^2)/2 +
     (t_8 T_8 \widetilde T_4^2)/2 + (T_{12} \widetilde T_4^3)/8\bigr)\cr
& -4 t_8^2 T_{14} - T_{12}^2 T_6 + 4 T_{10} T_{14} T_6 - 4 t_8 T_{14} T_2 T_6
+
  4 t_8 T_{14} T_8 - T_{14} T_8^2  - (T_{12}^2 T_2 \widetilde T_4)/2\cr
& +
  2 T_{10} T_{14} T_2 \widetilde T_4 - 2 t_8 T_{14} T_2^2 \widetilde T_4 - 2
t_8 T_{12} T_6 \widetilde T_4 -
  t_8 T_{14} \widetilde T_4^2 - t_8 T_{12} T_2 \widetilde T_4^2 - t_8^2 T_6
\widetilde T_4^2\cr
& + (T_{14} T_8 \widetilde T_4^2)/2 -
  (t_8^2 T_2 \widetilde T_4^3)/2 - (T_{14} \widetilde T_4^4)/16
}}
For the polynomial limit the $T_{2n}$ satisfy
$$T_{2n}=\sum_{i_i<i_2..<i_n}^8 m_{i_1}^2...m_{i_n}^2,\qquad\qquad
t_8=\prod_i^8 m_i,\qquad\qquad \widetilde T_4=T_2^2/4-T_4.$$
The expression differs slightly from the curve in \JMDN\ since we have
shifted the $x$ variable.

The lower $E_n$ curves are derived using the scaling described in the
text.
The $E_7$ curve, in terms of $SO(12)\times SU(2)$ variables is given
by
 \eqn\EVII{\eqalign{
y^2&=x^3+ \Bigl( u^2 - (u \wT_2)/2-2 t_6 - 2 t_6 T_2 - T_6  + 2 t_6 \wT_2
\Bigr) x^2\cr
& +
\Bigl ( u^3 (2 - T_2)- u^2(4 t_6 + T_4) + u\bigl (2 t_6 T_4-8 t_6 T_2 +
2 t_6 T_2^2
 - 2 T_8 +
     T_2 T_8 + 8 t_6 \wT_2 \cr
&- 3 t_6 T_2 \wT_2 + T_6 \wT_2 - T_8 \wT_2 +
 t_6 \wT_2^2/2\bigr)\cr
& +8 t_6^2 T_2 - 4 T_{10} T_2
 + T_{10} T_2^2 +
     2 t_6 T_2 T_6 + T_4 T_8  -
     8 t_6^2 \wT_2\cr
& + 4 T_{10} \wT_2 - 2 T_{10} T_2 \wT_2 + t_6 T_4 \wT_2 - 2 t_6 T_6 \wT_2 +
    T_{10} \wT_2^2 - (T_8 \wT_2^2)/4 - t_6 \wT_2^3/4 -
        8 t_6 \wT_2\cr
& - 3 t_6 T_2 \wT_2 + T_6 \wT_2 - T_8 \wT_2 + t_6 \wT_2^2/2\Bigr) x\cr
&+
   u^4( T_2^2/4-T_2) + u^3 (   2 t_6 T_2 -8 t_6+ T_4 \wT_2/2) +
  u^2\bigl(4 t_6^2 - 4 T_{10} + 8 t_6 T_4 - 3 t_6 T_2 T_4\cr
& + 2 T_2 T_8 -
     T_2^2 T_8/2 - 2 t_6 T_2 \wT_2 + t_6 T_2^2 \wT_2/2 + 2 t_6 T_4 \wT_2 -
     T_8 \wT_2 \cr
& + (T_2 T_8 \wT_2)/2 + t_6 \wT_2^2/2 - t_6 T_2 \wT_2^2/4 -
    (T_6 \wT_2^2)/4\bigr)\cr
& + u\Bigl(16 t_6^2 T_2 - 4 t_6^2 T_2^2 - 4 t_6^2 T_4
+
     4 T_{10} T_4 - 2 T_{10} T_2 T_4 - 2 t_6 T_4^2 + 8 t_6 T_2 T_6 -
     2 t_6 T_2^2 T_6 - 2 t_6 T_2 T_8 \cr
&- 16 t_6^2 \wT_2 + 6 t_6^2 T_2 \wT_2 -
     2 T_{10} T_2 \wT_2 + T_{10} T_2^2 \wT_2/2+
2 T_{10} T_4 \wT_2 - 8 t_6 T_6 \wT_2 +
     3 t_6 T_2 T_6 \wT_2 \cr
& + 2 t_6 T_8 \wT_2 - T_4 T_8 \wT_2/2 - t_6^2 \wT_2^2 +
    T_{10} \wT_2^2 - T_{10} T_2 \wT_2^2/2 + t_6 T_4 \wT_2^2/2 - t_6 T_6 \wT_2^2
+
     (T_8 \wT_2^3)/8\Bigr)\cr
 & -
8 t_6^3 T_2 + 8 t_6 T_{10} T_2 - 2 t_6 T_{10} T_2^2 + t_6^2 T_4^2 -
   T_{10} T_4^2 - 4 t_6^2 T_2 T_6 + 4 T_{10} T_2 T_6 - T_{10} T_2^2 T_6\cr
& -
   t_6 T_2 T_4 T_8 - T_2 T_8^2 + T_2^2 T_8^2/4  + 8 t_6^3 \wT_2
   - 8 t_6 T_{10} \wT_2 +
  4 t_6 T_{10} T_2 \wT_2\cr
& - t_6^2 T_2 T_4 \wT_2 + 4 t_6^2 T_6 \wT_2 - 4 T_{10} T_6 \wT_2 +
   2 T_{10} T_2 T_6 \wT_2 - 2 t_6 T_2 T_8 \wT_2 + t_6 T_2^2 T_8 \wT_2/2\cr
& +
  t_6 T_4 T_8 \wT_2 + T_8^2 \wT_2 - T_2 T_8^2 \wT_2/2 - 2 t_6 T_{10} \wT_2^2 -
  t_6^2 T_2 \wT_2^2 + t_6^2 T_2^2 \wT_2^2/4 + t_6^2 T_4 \wT_2^2/2\cr
& +
  T_{10} T_4 \wT_2^2/2 - T_{10} T_6 \wT_2^2 + 2 t_6 T_8 \wT_2^2 -
  3 t_6 T_2 T_8 \wT_2^2/4 + T_8^2 \wT_2^2/4 + t_6^2 \wT_2^3 -
  t_6^2 T_2 \wT_2^3/4  \cr
& + t_6 T_8 \wT_2^3/4 + t_6^2 \wT_2^4/16 -
  T_{10} \wT_2^4/16
}}
The $T_n$ variables have the same form as in the $E_8$ case, with
$T_{2n}=-T_{2n+2}+G_{2n}$ for $n>1$ and where
$$G_{2n}=\sum_{i_1<..i_n}^6 \sin^2m_{i_1}...sin^2m_{i_n}$$
We also have that
$$T_2=2\left(1-\prod_{i=1}^6 \cos m_i\right)\qquad\qquad
t_6=\prod_{i=1}^6\sin m_i$$
and
$$\wT_2=T_2-4\sin^2(\mu/2).$$
In deriving this curve, $x$ was shifted by $t_6T_2$.
In terms of the $E_7$ dimensions, where $[x]=6$, $[u]=4$ and $[T_n]=n$,
we see that all terms in the $E_7$ curve are either dimension 18 or
20.  The dimension 18 terms are what remain in the Kodaira limit.

The $E_6$ curve is
\eqn\Esixcurve{\eqalign{
y^2=&x^3+\Bigl( u^2+ u(2 \sin\lambda  - i T_2 e^{i\lambda})  + 4 i t_5-
T_4e^{2i\lambda} \Bigr)
x^2
\cr
&- e^{i\lambda}
\Bigl( 2 i u^3 + u^2\bigl(2 i \sin\lambda  T_2 +T_2e^{i\lambda}\bigr)+
  u\bigl (4 \sin\lambda  T_4e^{i\lambda}- 8 t_5 - 2 i T_6 + 4i\sin2\lambda t_5
  \bigr) \cr
&\qquad
+16 \sin^3\lambda  t_5 - 6 \sin\lambda  T_2 t_5 +
2 \sin\lambda  T_2 t_5e^{2i\lambda} + 4 \sin\lambda ^2 T_6e^{i\lambda} - T_2
T_6e^{i\lambda}
\cr
&\qquad+ 4 T_8 e^{-i\lambda}  -
  i t_5 (T_2^2-4T_4)e^{i\lambda}\Bigr)x\cr
&-
e^{2i\lambda}\Bigl(
  u^4 + 2 \sin\lambda  T_2 u^3 - u^2\bigl ( 2 T_6-4 \sin^2\lambda  T_4
- 4\sin2\lambda t_5\bigr)\cr
&\qquad -
  u\bigl ( 8 \sin\lambda  T_8 -8 \sin^3\lambda  T_6 + 2 \sin\lambda  T_2 T_6 -
4 \sin2\lambda\sin\lambda  T_2 t_5 + 2 \cos\lambda t_5(T_2^2-4T_4) \bigr)\cr
&\qquad
+ 4 \sin^22\lambda  t_5^2 + 4 \sin2\lambda t_5 T_6  +
 T_8(T_2^2-4T_4) + T_6^2
+ 16 \sin^4\lambda T_8 - 8 \sin^2\lambda  T_2 T_8 \Bigr).
}}
with
$$t_5=\prod_i^5 \sin m_i$$
The shift used on $x$ is $-it_5(2-T_2)e^{2i\lambda}+iT_2ue^{i\lambda}/2$.

The $E_5=SO(10)$ curve is
\eqn\Efivecurve{\eqalign{
y^2=& x^3 +
 \bigl ( u^2 + u(T_2 e^{i\lambda} - 4 \cos\lambda) - T_2 e^{2i\lambda} - 8 t_4
+
     T_2^2 e^{2i\lambda}/4\bigr) x^2\cr
& +
 \bigl (u^2( 4 - 2 T_2) + u( 4 T_2 e^{i\lambda}-8 t_4e^{-i\lambda}
 - T_2^2 e^{i\lambda} -
     4 T_4 e^{i\lambda})\cr
&  + 8i T_4\sin\lambda e^{i\lambda} + 16i t_4\sin\lambda( e^{i\lambda} -
2 e^{-i\lambda})
    + 12 t_4 T_2 + 16 T_6
- 2 T_2 T_4 e^{2i\lambda}) x\cr
& + u^2(T_2^2 - 4 T_2) + u(32i t_4\sin\lambda
+ 8 t_4 T_2e^{-i\lambda} -16i T_4\sin\lambda
 + 4 T_2 T_4 e^{i\lambda})\cr
&  - 16 t_4 T_4  - 16 T_2 T_6+
  16 t_4^2e^{-2i\lambda} +64 T_6\sin^2\lambda +
  4 T_4^2 e^{2i\lambda}
}}
In deriving this curve we shifted $x$ by
$(T_2-2)(ue^{i\lambda}/2 + t_4e^{2i\lambda})$.
If we set $T_n=t_4=\lambda=0$ then the curve in \Efivecurve\
reduces to
$$y^2=x^3+(u^2-4u)x^2+4u^2x$$
The discriminant of this curve is $128u^7-16u^8$, which describes
an $SO(10)$ singularity.  We could have also derived a curve where the
$SO(10)$ symmetry is  manifest by taking the curve in \Esixcurve\
and sending $\lambda$ to $i\infty$ and then rescaling.

The $E_4=SU(5)$ curve is
\eqn\Efourcurve{\eqalign{
y^2=&x^3+\bigl(u^2+u(iT_2e^{i\lambda}-2i(e^{i\lambda}+2e^{-i\lambda}))-
(1-T_2/2)^2e^{2i\lambda}-6T_2 - 16it_3+12\bigr)x^2\cr
&+\bigl(u(16iT_2 + 32t_3-32i-8ie^{2i\lambda})e^{-i\lambda}
-64T_4 + 16iT_2t_3+8T_2^2 - 32 i t_3\cr
&+4T_2(e^{2i\lambda}-8)+8(6-e^{2i\lambda})\bigr)-64iue^{-i\lambda} -
256t_3^2e^{-2i\lambda} + 256iT_2t_3e^{-2i\lambda}
-256T_4e^{-2i\lambda} + 128it_3-32T_2+16(4-e^{2i\lambda})}}
The shift in $x$ is
$(T_2-2)(it_3e^{2i\lambda}-2+iue^{i\lambda}/2)$.  In the
massless case, \Efourcurve\ reduces to
$$y^2=x^3+(u^2-6iu+11)x^2+(40-40iu)x+48-64iu$$
and the discriminant is $-256iu^5(u^2-iu+1)$ which describes an
$SU(5)$ singularity.

The $E_3=SU(3)\times SU(2)\times U(1)$ curve is
\eqn\Ethreecurve{\eqalign{
y^2=&x^3+\bigl(u^2+2u(e^{i\lambda}+4e^{-i\lambda}-T_2e^{i\lambda})
+12T_2 + 32t_2-24+\wT_4e^{2i\lambda}\bigr)x^2\cr
&64\bigl(u(T_2-2t_2-2)e^{-i\lambda}- 16t_2e^{-2i\lambda}-2\wT_4 + 2t_2T_2 -
2t_2+4
\bigr)\cr
&+
4096(t_2^2 - T_2t_2+\wT_4 + 2t_2-1)e^{-2i\lambda}
}}
where $\wT_4=(1-T_2/2)^2$.  The shift in $x$ is
$(T_2-2)(-t_2e^{2i\lambda}+1-ue^{i\lambda}/2)$.  In the massless case,
the curve reduces to
$$x^3+(u^2+10u-23)x^2+128(1-u)x$$
and the discriminant is
$-16384(1+u)^3(1-u)^2(17+u)$.  Hence this has an
$SU(3)\times SU(2)\times U(1)$
singularity structure.

The $E_2=SU(2)\times U(1)$ curve is
\eqn\Etwocurve{\eqalign{
y^2=&x^3+\bigl(u^2+iu(16e^{-i\lambda}+2e^{i\lambda} + t_1^2e^{i\lambda})
-\wT_4e^{2i\lambda}-24t_1^2 + 64it_1+48\bigr)\cr
&+256\bigl(u(2i + 2t_1-t_1^2)e^{-i\lambda}-2\wT_4 + 2it_1(1+16e^{-2i\lambda}) -
it_1^3\bigr)x\cr
&+65536(t_1^2-\wT_4 - it_1^3 + 2it_1)e^{-2i\lambda}}}
where $t_1=\sin m_1$ and $\wT_4=(1-sin^2m_1/2)^2$.  The shift in
$x$ is $(t_1^2-2)(-it_1e^{2i\lambda}-iue^{i\lambda}/2-8)$.
The massless curve is
$$x^3+(u^2+18iu+47)x^2+512(iu-1)x-65536$$
and its discriminant is
$4194394(5i+u)^2(-iu^3+23u^2-117iu-565)$.

The $E_1=SU(2)$ curve is
\eqn\Eonecurve{
y^2=x^3+\bigl(u^2-2u(e^{i\lambda}+16e^{-i\lambda})+e^{2i\lambda}-224\bigr)x^2-
65536e^{-2i\lambda}x}
The shift in $x$ is $-2e^{2i\lambda}-ue^{i\lambda}-32$.  In the
massless case, the discriminant is up to a numerical factor
$(u-17)^2(15+u)(u-49)$.

The $\widetilde E_1=U(1)$ curve is found by letting
$m_8=+i\sum\Lambda_i-2\lambda/8$, that is it has the opposite sign
as in the $E_1$ case.  This curve is given by
\eqn\Etonecurve{
y^2=x^3+\bigl(u^2-2u(e^{i\lambda}+16e^{-i\lambda})+32+e^{2i\lambda}\bigr)x^2+
4096\bigl(u e^{-i\lambda}-16e^{-2i\lambda}-1\bigr)x+4194304e^{-2i\lambda}
}

\goodbreak
\listrefs

\vfill
\eject
\end